\documentclass[fleqn,usenatbib]{mnras}

\usepackage[T1]{fontenc}
\usepackage{ae,aecompl}

\usepackage{graphicx}	
\usepackage{amsmath}	
\usepackage{amssymb}	
\usepackage{gensymb}
\usepackage{mathtools}
\usepackage[normalem]{ulem}
\usepackage{xcolor}

\usepackage{mathtools}
\usepackage{subcaption}
\usepackage{longtable}
\usepackage{pdflscape}

\usepackage{newtxtext,newtxmath}

\newcommand{\Rpar}{$\mathcal{R}$}
\newcommand{\CR}{$R_{cr}$}
\newcommand{\Om}{$\Omega_{bar}$}
\newcommand{\vint}{$\langle v \rangle$}
\newcommand{\xint}{$\langle x \rangle$}
\newcommand{\PAsym}{PA$_{sym}$}
\newcommand{\PAmod}{PA$_{mod}$}
\newcommand{\PAph}{PA$_{ph}$}

\newcommand{\quotes}[1]{``#1''}

\def\code#1{\texttt{#1}}
\def\Spear#1{$r_s={#1}$}

\title[Bar pattern speed in MWA galaxies]{SDSS IV MaNGA: Bar pattern speed in Milky Way Analogue galaxies}

\author[L. Garma-Oehmichen et al.]{
L. Garma-Oehmichen,$^{1}$\thanks{E-mail: lgarma@astro.unam.mx}
H. Hern\'andez-Toledo,$^{1}$
E. Aquino-Ort\'iz,$^{2}$
L. Martinez-Medina,$^{1}$
I. Puerari,$^{3}$
\newauthor
M. Cano-D\'iaz,$^{1}$
O. Valenzuela,$^{1}$
J. A. V\'azquez-Mata,$^{4, 5}$
T. G\'eron,$^{6}$
L. A. Martínez-Vázquez,$^{1}$ and
R. Lane$^{7}$\\
$^{1}$Instituto de Astronom\'ia, Universidad Nacional Aut\'onoma de M\'exico, Apartado Postal 70-264, CDMX, 04510, M\'exico  \\
$^{2}$Instituto de Astrof\'isica, Pontificia Universidad Cat\'olica de Chile, Av. Vicuña Mackenna 4860, 782-0436 Macul, Santiago, Chile.\\
$^{3}$Instituto Nacional de Astrof\'\i sica, Optica y Electr\'onica, Apdo. Postal 51 y 216, 72000 Puebla, Puebla, M\'exico \\
$^{4}$Departamento de F\'isica, Facultad de Ciencias, Universidad Nacional Aut\'onoma de M\'exico, Ciudad Universitaria, CDMX, 04510, M\'exico\\
$^{5}$Instituto de Astronom\'ia sede Ensenada, Universidad Nacional Aut\'onoma de M\'exico, Km 107, Carret. Tij.-Ens., Ensenada, 22060, BC, M\'exico\\
$^{6}$Oxford Astrophysics, Department of Physics, University of Oxford, Denys Wilkinson Building, Keble Road, Oxford, OX1 3RH, UK. \\
$^{7}$ Centro de Investigaci\'on en Astronom\'ia, Universidad Bernardo O'Higgins, Avenida Viel 1497, Santiago, Chile}
\date{Accepted XXX. Received YYY; in original form ZZZ}

\pubyear{2022}

\begin{document}
\label{firstpage}
\pagerange{\pageref{firstpage}--\pageref{lastpage}}
\maketitle

\begin{abstract}
Most secular effects produced by stellar bars strongly depend on the pattern speed. Unfortunately, it is also the most difficult observational parameter to estimate. In this work, we measured the bar pattern speed of 97 Milky-Way Analogue galaxies from the MaNGA survey using the Tremaine-Weinberg method. The sample was selected by constraining the stellar mass and morphological type. We improve our measurements by weighting three independent estimates of the disc position angle.
To recover the disc rotation curve, we fit a kinematic model to the H$_\alpha$ velocity maps correcting for the non-circular motions produced by the bar. The complete sample has a smooth distribution of the bar pattern speed  ($\Omega_{Bar}=28.14^{+12.30}_{-9.55}$ km s$^{-1}$ kpc $^{-1}$), corotation radius ($R_{CR} = 7.82^{+3.99}_{-2.96}$ kpc) and the rotation rate ($\mathcal{R} = 1.35^{+0.60}_{-0.40}$). 
We found two sets of correlations: (i) between the bar pattern speed, the bar length and the logarithmic stellar mass (ii) between the bar pattern speed, the disc circular velocity and the bar rotation rate.
If we constrain our sample by inclination within $30 \degree < i < 60 \degree$ and relative orientation $20\degree<|PA_{disc}-PA_{bar} |<70\degree$, the correlations become stronger and the fraction of ultra-fast bars is reduced from 20\% to 10\% of the sample.
This suggest that a significant fraction of ultra-fast bars in our sample could be associated to the geometric limitations of the TW-method.
By further constraining the bar size and disc circular velocity, we obtain a sub-sample of 25 Milky-Way analogues galaxies with distributions $\Omega_{Bar}=30.48^{+10.94}_{-6.57}$ km s$^{-1}$ kpc$^{-1}$, $R_{CR} = 6.77^{+2.32}_{-1.91}$ kpc and $\mathcal{R} = 1.45^{+0.57}_{-0.43}$, in good agreement with the current estimations for our Galaxy.

\end{abstract}

\begin{keywords}
galaxies: disc -- galaxies: evolution -- galaxies: kinematics and dynamics -- galaxies: structure
\end{keywords}



\section{Introduction}
\label{sec:introduction}

Stellar bars exist in a great variety of shapes, sizes, and galactic environments. Most of their properties are strongly tied to the stellar mass and morphology of their host galaxy. For instance, the bar fraction (the likelihood of hosting a large-scale bar) and bar length are strongly dependent on the galaxy mass \citep{Nair2010, Masters2012, Erwin2018}. Early-type galaxies host stronger bars than their late-types counterparts. They are larger in relation to their disc \citep{MendezAbreu2012, Diaz2016_Bar_characteristics, Erwin2018}, prolate shaped \citep{Diaz2016_Bar_characteristics, Mendez-Abreu2018} and have a flat density profile \citep{Elmegreen1985, Kim2015}. Moreover, the size ratio between the bar and the disc remains constant over the cosmic time, suggesting an efficient coupling between both structures \citep{Perez2012, Kim2021}.

Bars are one of the main drivers of the galaxy secular evolution \citep{Weinberg1985, Kormendy2004, Sellwood2014, Diaz-Garcia2016_Secular_Evolution}. Numerous analytical and numerical studies show they are efficient at transferring angular momentum from their inner resonances to those outside of corotation via dynamical friction \citep{LyndenBell1972, Tremaine1984b, Athanassoula2003}. This angular momentum is  mostly absorbed by the dark matter halo \citep{Weinberg1985, Debattista2000}, and in a smaller fraction by the bulge \citep{Kataria2019}.

As a result, the bar is expected to induce substantial gaseous flows to the galaxy centres \citep{Sormani2015}. Observations of barred galaxies show a clear increase in the concentration of molecular gas \citep{Sakamoto1999, Jogee2005} and in the star formation rate \citep{Ellison2011, Chown2019} in the central regions. 
If the process in not balanced with the inflow of cosmological gas, the bar can deplete the gas supply in the disc, causing the so-called \quotes{bar quenching} \citep{Masters2012}. This scenario is supported by  observations of the specific star formation rate  \citep{Cheung2013}, colour \citep{Gavazzi2015, Kruk2018}, gas fraction, \citep{Newnham2020}, star formation histories \citep{Fraser2020} and statistical properties of the galaxies \citep{Geron2021}.

A theoretical consequence of the large scale gas flows is the flattening of the metallicity profile \citep{Cavichia2014, Kubryk2015Abundance_profiles}. However, this behaviour does not appear in most observations \citep{Sanchez-Blazquez2014, Perez-Montero2016, Sanchez-Menguiano2016}, and has only been observed in low luminosity (low mass) galaxies \citep{Zurita2021}.

The bar is mostly supported by regular resonant stellar orbits located inside corotation. The most important being the x1 family \citep{Contopoulos1980}. Nonetheless, chaotic orbits are also important building blocks for the bar. For example, by modelling N-body simulation snapshots, various authors have shown that regular and sticky orbits can eventually transform to chaotic orbits that support the X-shaped/boxy structure \citep{Voglis2007, Harsoula2009, Chaves-Velasquez2017}. Moreover, chaotic orbits near the Lagrangian points could be responsible for the support of the spiral structure according to the Manifold theory \citep{Voglis2006, Romero2006, Romero2015}. The strong correlation between the strengths of the bar and spiral arms suggest both structures could be intimately coupled \citep{Salo2010, Diaz2019, Garma-Oehmichen2021}.

The multiple resonances produced by bars and spiral arms also have a profound effect on the disc stellar orbits. Simulations show that near corotation, stars can scatter inwards or outwards without changing their orbital ellipticity in a process called \quotes{radial migration}  \citep{Sellwood2002, Kubryk2015radial_migration}. Since stars preserve their circular orbits, this process does not contribute to the radial heating of the disc, but should affect the metallicity distribution \citep{Martinez-Medina2017}. The rearrangement of stars can also trigger the creation of moving groups \citep{Perez-Villegas2017} and resonant trapping stars in the disc and the stellar halo \citep{Quillen2014, Moreno_etal2015}.

All these secular effects depend in great extent on the bar pattern speed (hereafter \Om{}). Unfortunately, it is the most difficult observational parameter to estimate. Most methods developed to measure \Om{} require some modelling. Some use the gas flow induced by the bar, for example, by matching observations with hydrodynamical simulations \citep{Sanders1980, Hunter1988, England1990, Weiner2001, Perez2004, Zanmar-Sanchez2008, Rautiainen2008}, or studying the residual gas velocity field after subtracting a rotation model \citep{Font2011, Font2014, Font2017}. Other methods are based on the location and shape of different morphological features like dark gaps in ringed galaxies \citep{Buta2017_DarkGaps, Krishnarao2022}, the offset of dust lanes \citep{Athanassoula1992, Sanchez-Menguiano2015}, changes in the phase of spirals \citep{Puerari1997, Aguerri1998, Sierra2015} or the position of rings \citep{Buta1986, Rautiainen2000, Patsis2003}.

The only direct method for estimating \Om{} is the so-called \cite{Tremaine1984} (hereafter TW) method (see Section \ref{sec:TW}). It requires the surface brightness and line-of-sight (LOS) velocity of a tracer that suffices the continuity equation. Until recently, most measurement using the TW-method were made with long-slit spectroscopy in early-type galaxies \citep[e.g.][]{Kent1987, Merrifield1995, Gerssen1999, Debattista2002, Aguerri2003, Corsini2003, Debattista2004, Corsini2007}.

With the advent of the integral field spectroscopy technique, the TW method has been applied to an increasing number of galaxies. \cite{Aguerri2015} measured \Om{} in 15 strong barred galaxies from the survey CALIFA and found no trend with the morphological type. \cite{Guo2019} used a sample of 53 galaxies from the MaNGA survey and studied the effects of the galaxy position angle and inclination on the determination of \Om{} (see also \cite{Debattista2003}). \cite{Cuomo2019} continued the measurements from \cite{Aguerri2015} within the CALIFA survey, finding weakly barred galaxies have similar values of \Om{} as the strongly barred galaxies. In \cite{Garma-Oehmichen2020} (hereafter G20), we used a sample of 15 MaNGA galaxies and 3 CALIFA galaxies to study different uncertainty sources and identify systematic errors in the method. \cite{Williams2021} applied the TW-method to 19 galaxies from the PHANGS-MUSE survey. They found ISM tracers produce erroneous signals in the TW integrals because of their clumpy distribution.

The rotation rate parameter $\mathcal{R} = R_{CR}/R_{Bar}$ (the ratio between corotation and the bar radius) gives a dynamical interpretation to the bar. The physical significance of this ratio comes from the fact that the stellar orbits that support the bar cannot extend beyond corotation \citep{Contopoulos1980, Athanassoula1980}. Thus, the faster the bar rotates in relation to the disc, the closer \Rpar{} goes to 1. This ratio is commonly used to classify bars as slow (\Rpar{} > 1.4), fast (1 < \Rpar{} < 1.4) and the theoretically un-physical ultra-fast (\Rpar{} < 1).

Recent measurements using the TW-method find that most bars are consistent with a fast classification (with a concerning amount of ultra-fast bars) \citep{Guo2019}. This has led to a growing tension with the $\Lambda$CDM cosmological framework, where bars slow down excessively because of dynamical friction with the dark matter halo \citep{Algorry2017, Peschken2019, Roshan2021_mond, Roshan2021}. Nonetheless, \cite{Fragkoudi2021} showed that the bars can remain fast in more baryon dominated galaxies with higher stellar-to-dark matter ratios. 

The properties of the dark matter halo are incredibly important for the bar formation and evolution.  Several N-body simulations have shown that more concentrated halos are able to form stronger and larger bars \citep{Debattista1998, Debattista2000, Athanassoula2002}. Triaxial halos can induce the formation of bars \citep{Valenzuela2014} but produce weaker bars compared to spherical halos \citep{Athanassoula2013}. Recent works have shown the importance of the halo spin. For example, a spinning halo can effectively suppress bar formation, by being unable to absorb angular momentum with the same efficiency \citep{Long2014, Collier2018, Rosas-Guevara2022}. Spinning halos can also change the radial extent of the disc \citep{Grand2017}, affecting the bar instability criteria \citep{Izquierdo-Villalba2022}.

Several attempts have been made in the last decade to estimate \Om{} in the Milky Way (MW). Arguably, the best estimations have been derived using the made-to-measure method, which matches the density of red clump giants with N-body models of barred galaxies \citep[][]{Portail2017, Perez-Villegas2017, Clarke2019, Clarke2022}. Other attempts have used an adaptation of the TW method \citep{Bovy2019}, or kinematic maps of disc and bulge stars, along with the continuity equation \citep{Sanders2019}. Other methods are based on N-body simulations of MW-like galaxies \citep{Tepper-Garcia2021, Kawata2021}. All these measurements seem to converge to a pattern speed of $\Omega_{bar}\sim$ 35 - 40 km s$^{-1}$ kpc$^{-1}$. This value corresponds to a corotation radius $R_{cr} \sim 5-6$ kpc, and $\mathcal{R} \sim 1.2$, therefore classifying the MW bar in the fast category.  

In this paper, we tie together the bar pattern speed of our galaxy in the context of extra-galactic measurements. We choose a sample of galaxies that reassemble the Milky Way in morphology and stellar mass. In G20 we argued some uncertainties in the TW method were still not well understood (see also Discussion in \cite{Guo2019}), and many improvements could be made. In line with this, we weight various independent measurements of the disc position angle (PA), we changed the distributions for sampling various parameters and change the rotation curve model. Throughout our procedure, we highlight possible systematic errors and decisions that could be biasing our measurements. All together, this has made our measurements more robust.

We organised the paper as follows: In Section \ref{sec:data} we introduce the MaNGA survey and the data processing packages used. Section \ref{sec:TW} gives an overview of the Tremaine-Weinberg method, and discuss its limitations. In Section \ref{sec:sample} we present our sample selection. In Section \ref{sec:meassurements} we show our measurement procedures and discuss the error estimation. We present the complete sample results in section \ref{sec:results} and discuss future improvements Section \ref{sec:discussion}.  Finally in Section \ref{sec:conclusions} we summarise our conclusions. All Figures and relevant measurements are available at the public repository: \url{https://github.com/lgarma/MWA_pattern_speed}
\section{Data}
\label{sec:data}



Mapping Nearby Galaxies at Apache Point Observatory (MaNGA) \citep{Bundy2015} is the largest galaxy survey using the Integral Field Spectroscopy (IFS) technique. It observed a sample of over $10,000$ galaxies. Its main sample comprises galaxies of all morphological types \citep[e.g.][]{VazquezMata2022}, within the redshift interval $0.1 < z <0.15$ and stellar masses between $10^{9}$ and $10^{12}$ $M_{\odot}$ \citep[for details about the sample see:][]{Wake17}. The final sample also includes objects selected as part of $20$ ancillary programs driven by specific scientific objectives \citep[for details refer to:][]{Abdurro2021}, so a small fraction contains galaxies that may lie outside the aforementioned characteristics.

The MaNGA survey is one of the three major programs of the SDSS-IV international collaboration \citep{Blanton2017}. Its observations are carried out by the Sloan 2.5 meter telescope \citep{Gunn2006}, and uses the BOSS spectrograph, which provides a spectral resolution of $R \sim 2000$ in the wave-length range $3600$ {\AA} - $10300$ {\AA} \citep{Smee2013}.

Its Integral Field Units (IFUs) consist of a set of 17 hexagonal fiber-bundles of five different sizes ranging from 19 to 127 fibers \citep{Dory2015}. 
Each fiber has a diameter of 2 arcsec. Data taking requires that these fiber bundles be plugged into previously drilled plates, whose positions correspond to the targets to be observed per night. Finally, a 3-point dithering observing strategy is used to ensure total coverage of the required field of view \citep{Law2016}, which for the MaNGA targets vary from $\sim$1.5 to $\sim$2.5 $R_{e}$.

In this work, we use the MaNGA reduced data \citep[see][for details]{Law16} as well as the following data products provided by the Data Analysis Pipeline (DAP) \citep{Westfall2019}: Mean flux in r-band (as a proxy of the stellar flux), stellar velocity, the H$_\alpha$ velocity and their respective dispersion maps. The data was accessed via the Python package \code{Marvin}  \citep{Cherinka2019}. All DAP datacubes are Voronoi-binned based on the g-band weighted signal-to-noise ratio (S/N) with a target S/N of 10 per Voronoi bin. The reconstruction method minimises the spatial covariance, making the number of independent measurements similar to the number of fibers \citep{Liu2020}.

Additionally, we use \code{Pipe3D} to recover some integrated properties, including the stellar mass, molecular gas, stellar spin parameter, stellar surface density and the velocity-sigma ratio. \citep{Sebastian2015, Sebastian2016, Sanchez2018}.

We will refer to the galaxies using their MaNGA identification number, which consist of two sets of numbers (for example 8596-12704). The first set (8596) corresponds to plate id, while the second set combines the number of fibers (127) and the fiber-bundle used (04).

\section{Tremaine Weinberg method}
\label{sec:TW}




Using a tracer that follows the continuity equation and assuming a perfectly flat disc, the \cite{Tremaine1984} method estimates \Om{} as the ratio between two integrals:

\begin{equation}
    \Omega_{bar} \sin i = \frac{\langle v \rangle }{\langle x\rangle}
\end{equation}

where $i$ is the galaxy inclination, and \vint{} and \xint{} are the intensity-weighted means of the line-of-sight velocity and position of the tracer respectively \citep{Merrifield1995}. In equations this is:

\begin{equation}
    \langle v  \rangle   = \frac{\int_{-\infty}^{+\infty} h(y) dy \int_{-\infty}^{+\infty}V_{LOS}(x,y) \Sigma(x,y)  dx }{\int_{-\infty}^{+\infty} h(y) dy \int_{-\infty}^{+\infty}\Sigma(x,y) dx}
\end{equation}

\begin{equation}
      \langle x  \rangle  = \frac{\int_{-\infty}^{+\infty} h(y) dy \int_{-\infty}^{+\infty} x \Sigma(x,y) dx  }{\int_{-\infty}^{+\infty} h(y) dy \int_{-\infty}^{+\infty}\Sigma(x,y) dx}  
\end{equation}

where $(x,y)$ is the Cartesian coordinate system of the sky plane, with $x$ aligned with the line of nodes (i.e the disc position angle). $\Sigma$ is the surface brightness and $V_{LOS}$ is the velocity in the line-of-sight, of the tracer. In this work we are using the old stellar component for weighting the TW integrals, with the r-band mean flux datacube for $\Sigma$ and the stellar velocity map for $V_{LOS}$ \citep{Westfall2019}.

$h(y)$ is an arbitrary odd weight function ($h(-y) = -h(y)$). To simulate the position of $2m+1$ slits offset by a separation $ y_0$ we use a sum of Dirac deltas $h(y) = \sum_{n=0}^{m}\delta(y - n y_0) - \delta(y + ny_0)$

The separation between slits $y_0$ should be wide enough to prevent the use of repeated information. This separation depends on the relative orientation between the slits and the Cartesian grid:

\begin{equation}
    y0 = p / \cos(\theta)
\end{equation}

where $p$ is the pixel separation and $\theta$ is lowest the angle between the grid and the disc position angle (see equation 6 in G20). We choose to set the pixel separation to $p=1$ between consecutive slits.

The symmetry in $x$ and $y$ in both \vint{} and \xint{}, makes all axisymmetric features to cancel out in the integration. Thus, in practice, the integration limits can be set to cover only the non-axisymmetric structure. In some galaxies, however, the IFU coverage is not enough to reach the axisymmetric disc. It is possible a systematic error could be happening in those galaxies. We cannot estimate the value of \Om{} using a larger spatial coverage, but we can reduce the length of the slits to do the opposite. To get an estimate of this systematic error, we include a slits length error in our measurements.

\begin{figure*}
    \centering
    \includegraphics[width=\linewidth]{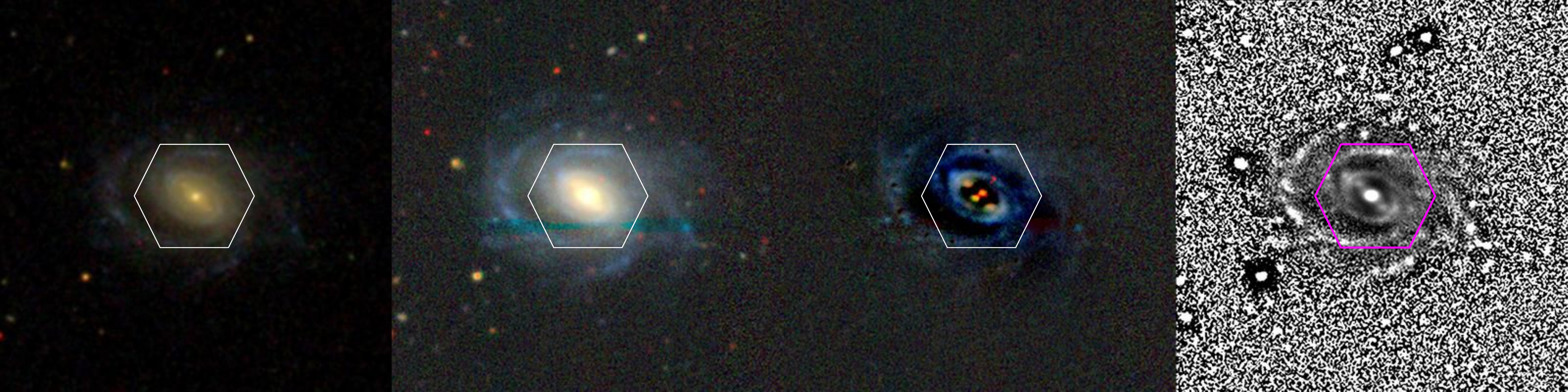}
    \caption{Image processing of the example galaxy 8596-12704. From left to right: (i) SDSS $gri$ composite colour image, (ii) DESI $grz$ composite colour image, (iii) DESI Residual image after subtracting the best Bulge/Disc 2D model of the galaxy brightness, (iv) Filter-enhanced DESI r-band image. All images include the MaNGA FoV.}
    \label{fig:mosaic}
\end{figure*}

\begin{figure*}
    \centering
    \begin{subfigure}{0.33 \textwidth} 
    \includegraphics[width=\textwidth]{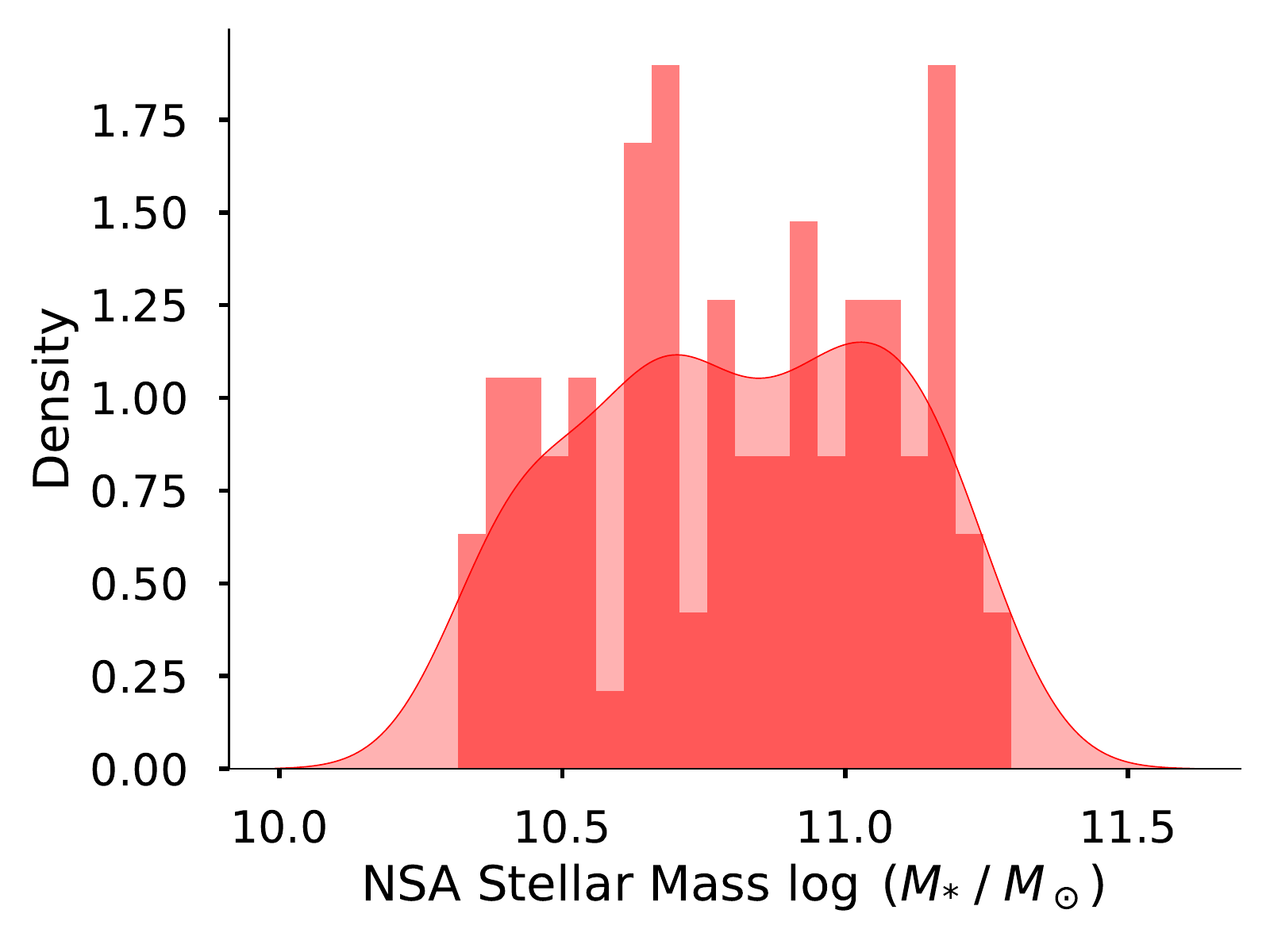}
    \end{subfigure}
    \begin{subfigure}{0.33 \textwidth} 
    \includegraphics[width=\textwidth]{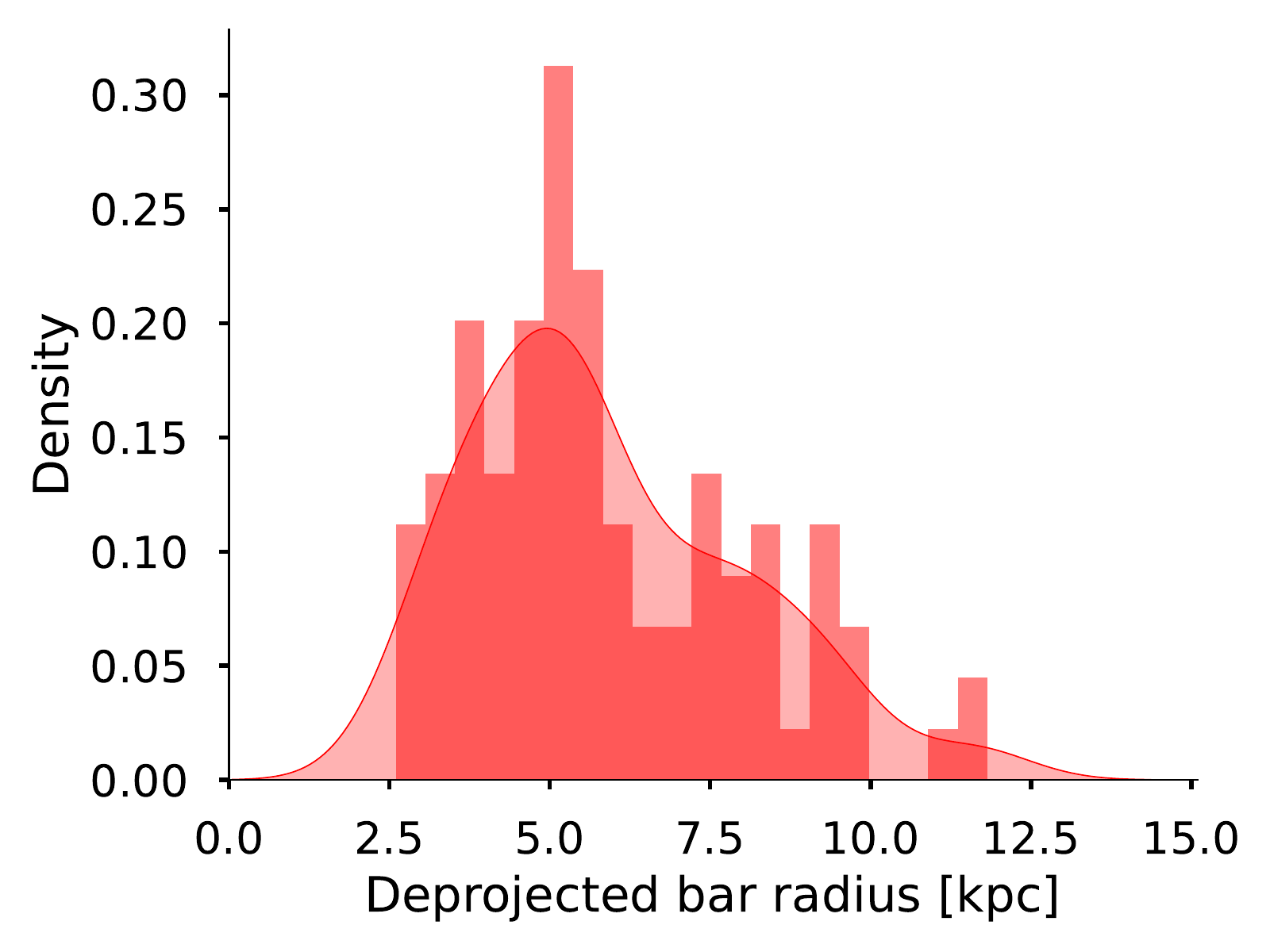}
    \end{subfigure}
    \begin{subfigure}{0.33 \textwidth} 
    \includegraphics[width=\textwidth]{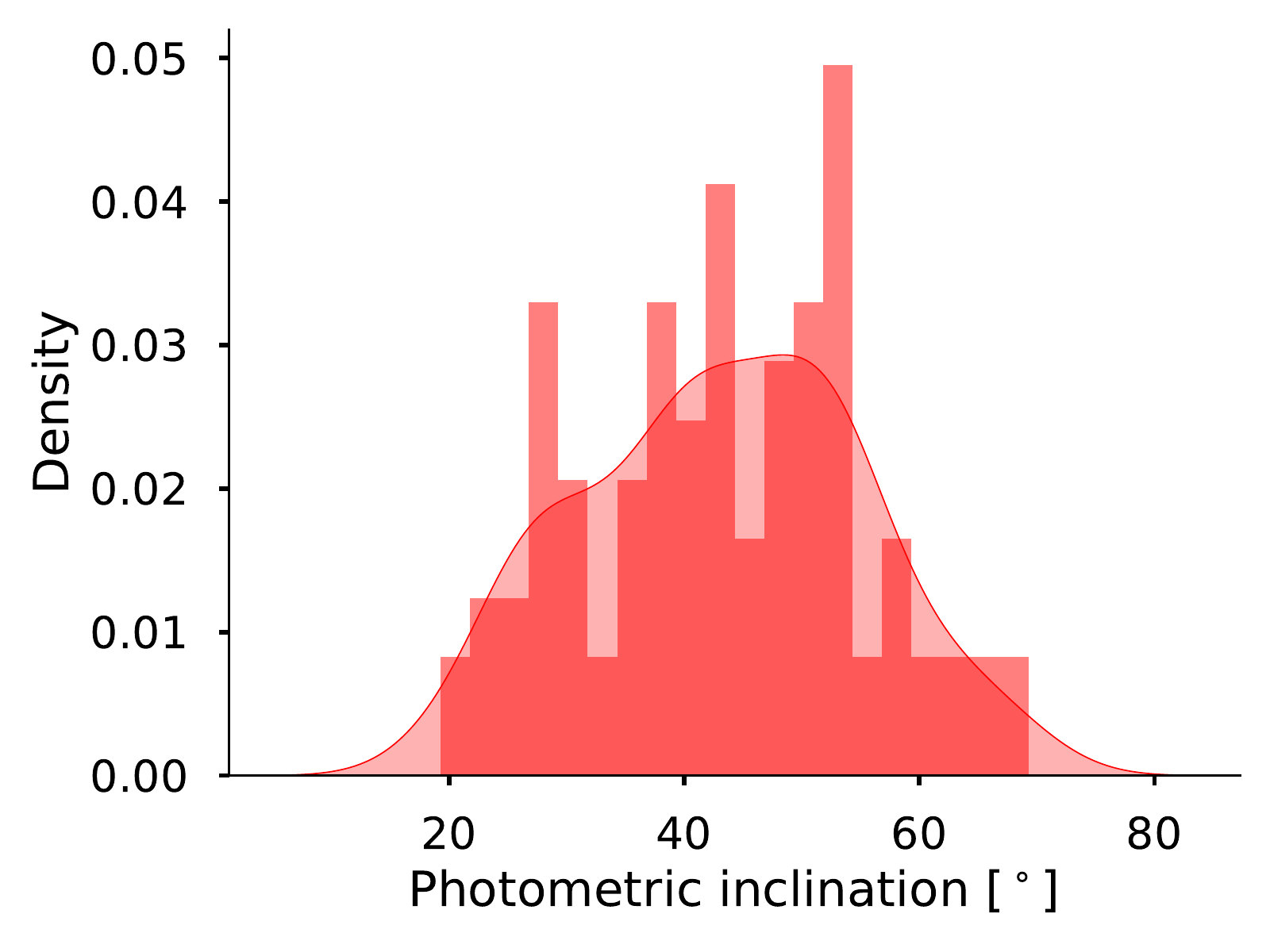}
    \end{subfigure}
    \caption{Statistical distribution of our sample. A kernel density estimate is used to smooth the distribution. \textit{Left panel}: Logarithmic stellar mass from the NASA-Sloan Atlas catalogue. \textit{Middle panel}: Deprojected bar radius, estimated from an isophotal analysis. \textit{Right panel}: Disc inclination, estimated from the isophotal analysis.}
    \label{fig:sample_statistics}
\end{figure*}

The symmetry property also makes the TW method extremely sensible to the disc PA, as a wrongful estimation introduces a false signal to the integrals \citep{Debattista2003}. It is common that different estimates of the disc PA are not consistent within their uncertainties. This introduces another systematic error that is heavily dependent on the author's criteria. In G20, we compared the measurements of 10 galaxies in common with \cite{Guo2019} and discussed differences in procedures to estimate \Om{}. The average difference in the disc PA was 3$\degree$, enough to change most measurements significantly. In G20, our measurements of \Om{} differed on average 10 km s$^{-1}$ kpc$^{-1}$ when using a photometric or kinematic disc position angle. In this work, we weight the PA of three independent measurements (see section \ref{sec:PA_measure}).

The TW method cannot be applied to all barred galaxies. If the bar is oriented towards the disc major or minor axis, the TW integrals will become symmetric and cancel the non-axisymmetric contributions. Thus, galaxies with mid inclinations are preferable.
\section{Sample selection}
\label{sec:sample}



By assuming that the Milky Way is not extraordinary among its peers, we can circumvent the limitation of observing our galaxy by studying galaxies with similar properties in the extra-galactic context \citep{Boardman2020_Andromeda}. Throughout the literature, there have been many definitions of what makes a good Milky Way Analogue (MWA), usually based on the science goals of the study. Common limits include the stellar mass, the bulge-to-total ratio \citep{Boardman2020}, the star formation rate \citep{Licquia2015}, galactic companions \citep{Robotham2012, Boardman2020_Andromeda} and the presence of galactic structures like the bar, spiral arms \citep{Fraser-McKelvie2019} or an X-shaped pseudo-bulge \citep{Georgiev2019, Kormendy2019}.

Due to the nature of this work, we tailor our selection criteria to the stellar mass and the Hubble morphological type. The MW most prominent features include a strong peanut-shaped bar and four major spiral arms. Thus, its Hubble classification may lie somewhere between SBb and SBc \citep{Shen2020}. We used the visual morphological classification reported in \cite{VazquezMata2022}. Figure \ref{fig:mosaic} shows a mosaic containing a set of post-processed images from the SDSS and DESI Legacy Image Surveys used for morphological evaluation.

The stellar mass of the MW is estimated to be $log(M_* / M_\odot) = 10.75 \pm 0.2$ \citep{Licquia2016}. We choose to extend this range to cover a complete dex in stellar mass $\log M \sim [10.3 - 11.3]$. Stellar masses come from a cross-match with the NASA-Sloan Atlas (NSA) catalogue \citep{Blanton2011}, using $h=0.71$ Hubble constant and a Chabrier initial mass function \citep{Chabrier2003}. 

Using these cuts, we end up with a sample of 233 galaxies. Still, not all barred galaxies are good candidates for the TW method. The TW methods weights both the kinematic and positional data, so the inclination becomes an important selection parameter. Face-on galaxies tend to have poor kinematic data but great positional data. The opposite occurs in edge-on galaxies. To filter these galaxies, we include cuts in inclinations within the range $20 \degree < i < 70 \degree$. 

As mentioned in Section \ref{sec:TW}, the bar and the disc should not be oriented parallel nor perpendicular to each other. We included an additional cut in their relative orientation: $10 \degree < |PA_{Bar} - PA_{Disc}| < 80 \degree$.

Finally, a good measurement with the TW method should produce a clear linear relation in between both TW-integrals. However, some galaxies do not display this trend, possibly because a lack of symmetry. We choose to remove those galaxies as well.

All together, we discarded $\sim 60 \%$ of the galaxies, bringing our final sample to 97 MWA galaxies. In Figure \ref{fig:sample_statistics} we show the stellar mass, bar radius and disc inclination distributions of the final sample. In Section \ref{sec:mw_like} we discuss a sub-sample where we further increase the selection criteria to include the bar size and disc circular velocity to be more similar to the MW. Divided by the number of fibers, our sample contains 1, 14, 18, 12 and 52 galaxies with 19, 37, 61, 91 and 127 fibers respectively.

\section{Measurements}
\label{sec:meassurements}

Throughout this section, we will present our procedure using the galaxy 8596-12704  shown in Figure \ref{fig:mosaic}. This galaxy possesses a strong bar with an inner ring near the end of the bar. It is a grand design galaxy with two strong symmetrical arms that extend to the outer region. The MaNGA field of view only covers the bar and the inner ring.

\begin{figure}
    \centering
    \includegraphics[width=\linewidth]{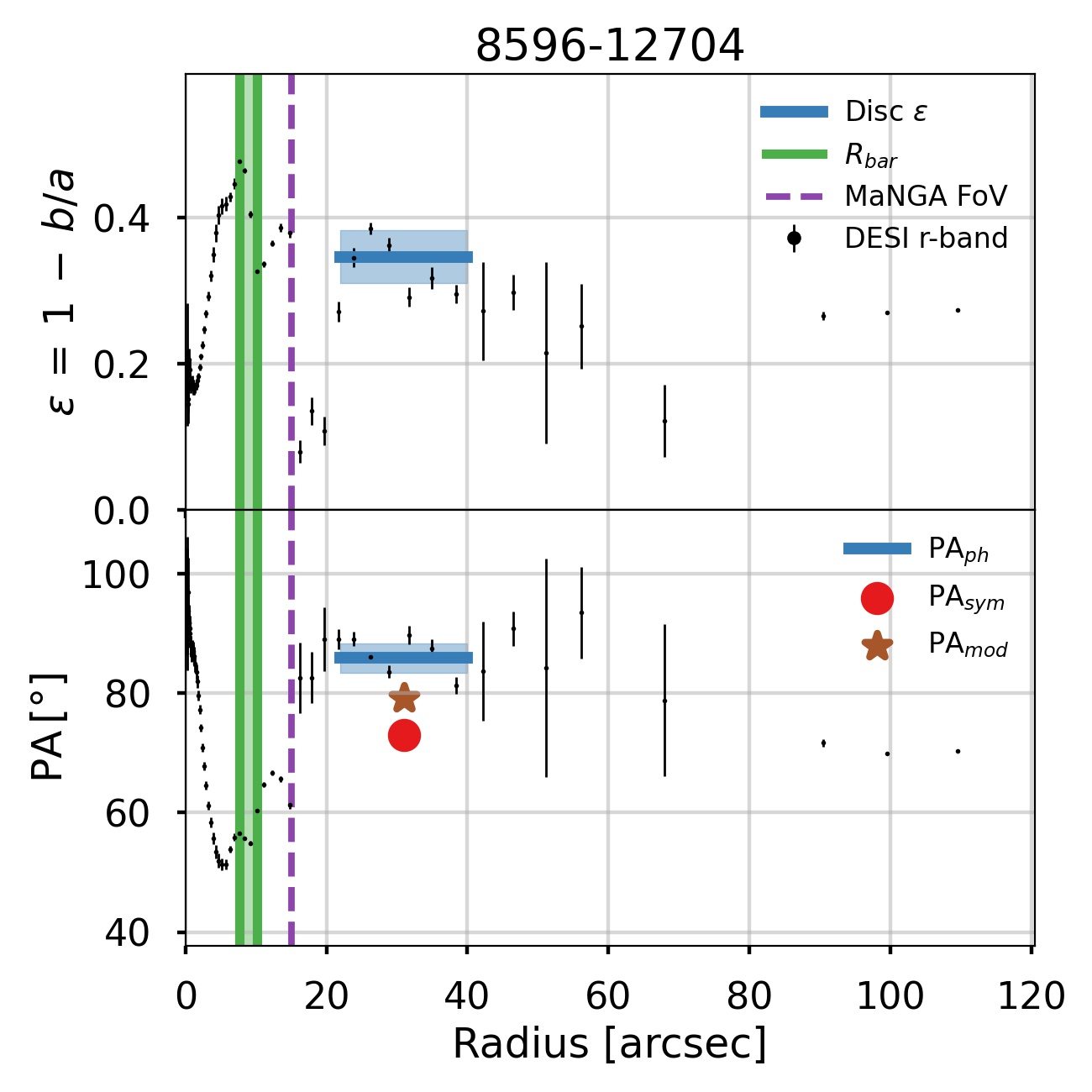}
    \caption{Ellipticity and position angle profiles of 8596-12704. The bar radius is shown with a green vertical stripe, limited by $R_\epsilon$ and $R_{PA}$. A sudden jump in ellipticity near the bar end occurs because of the inner ring. We estimate the disc PA using the isophotes between 22 and 40 (arcsec) shown as a blue region. For comparison, we also present the estimates after symmetrizing the stellar velocity field (PA$_{sym}$) and from the H$_\alpha$ kinematic model (PA$_{mod}$). In this galaxy, the three measurements of PA do not agree within uncertainties.}
    \label{fig:isophotes}
\end{figure}

In Figure \ref{fig:isophotes} we show the corresponding isophotal profiles performed over the DESI r-band images \citep{Dey2019}. We used the ELLIPSE routine \citep{Jedrzejewski1987} within the IRAF environment \citep{Tody1993}. To increase the signal of the outer disc, we sample the isophotes using a logarithmic step with the centre fixed. 

The isophotes provide a photometric estimate of the disc PA, inclination, bar radius and bar PA. We discuss different biases in these measurements in sections \ref{sec:PA_measure} and \ref{sec:bar_radius}. 

\begin{figure*}
    \centering
    \begin{subfigure}{0.5 \textwidth}  
    \includegraphics[width=\textwidth]{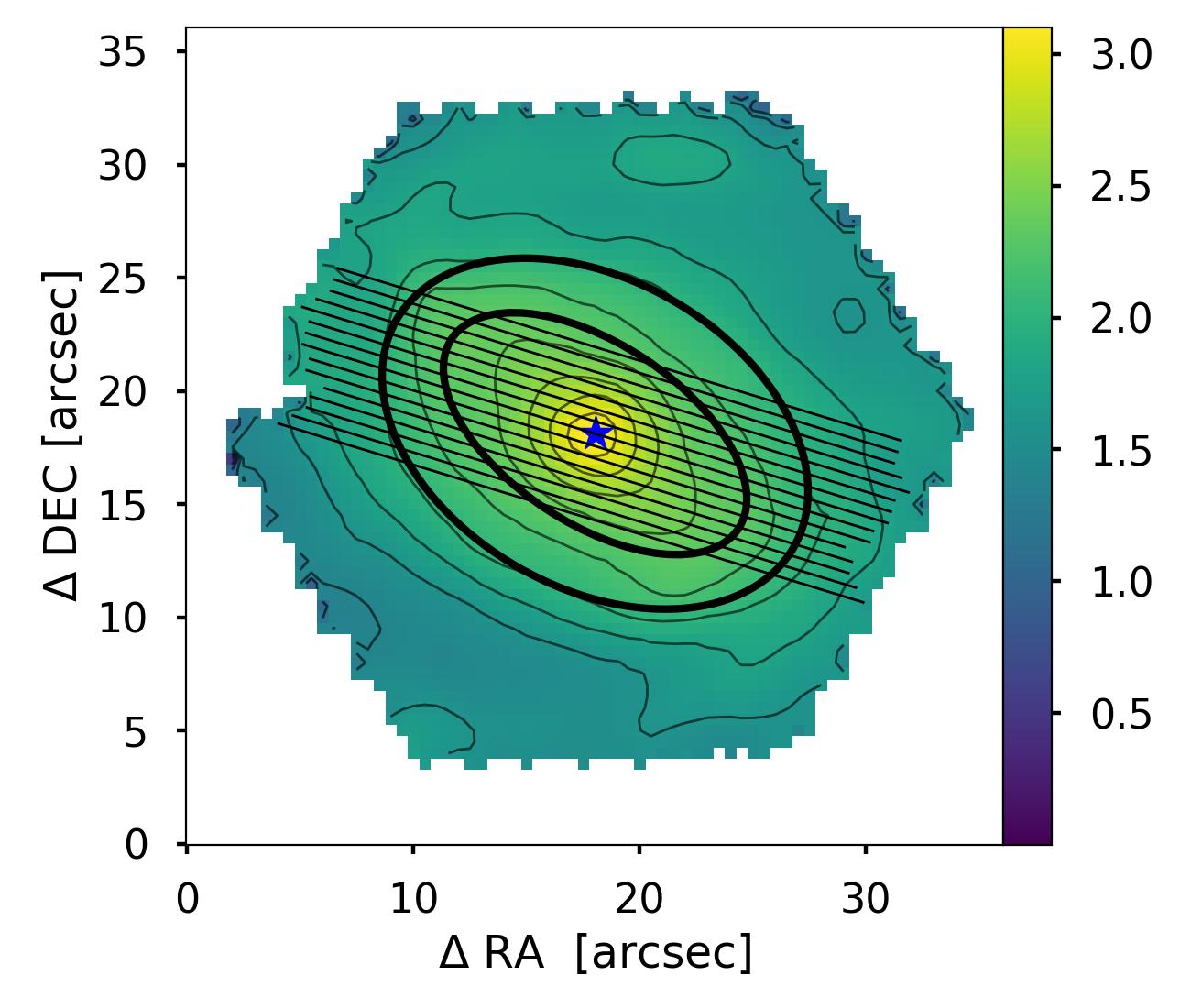}
    \end{subfigure}%
    \begin{subfigure}{0.5 \textwidth}  
    \includegraphics[width=\textwidth]{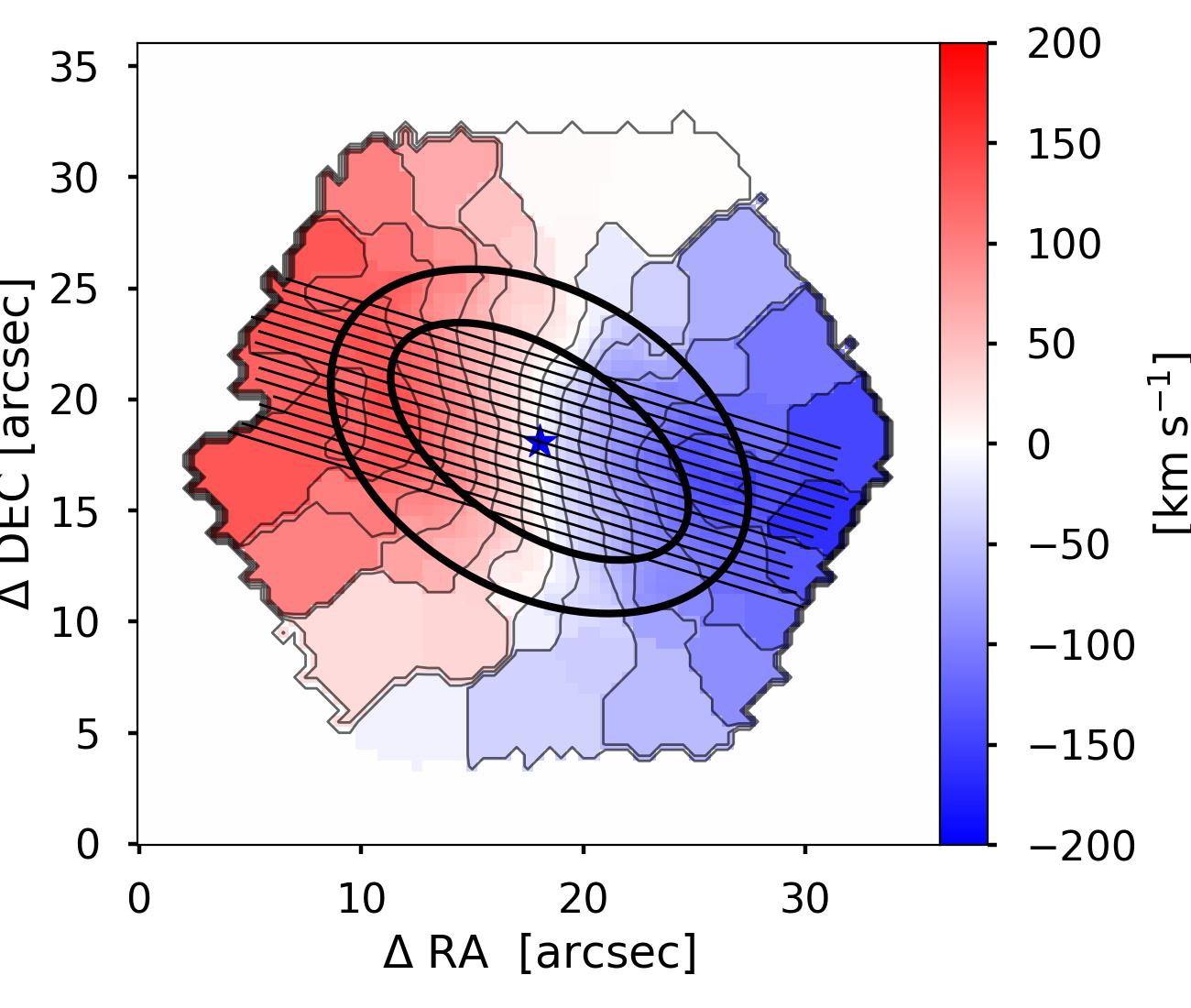}
    \end{subfigure}
    \caption{The r-band surface brightness and stellar velocity maps of 8596-12704. We display the surface brightness with a logarithmic scale for visualisation purposes. The bar estimates from the isophotal profiles are shown as black ellipses. We compute the TW integrals over the slits that cover the barred region.}
    \label{fig:flux_vel}
\end{figure*}

In Figure \ref{fig:flux_vel} we show the r-band surface brightness and stellar velocity maps, obtained from \code{Marvin}. We compute the TW integrals over these two maps.

\subsection{Disc Position Angle}
\label{sec:PA_measure}

The disc PA is the most important parameter for making reliable measurements of \Om{} with the TW method. It is well documented that few degrees of error in PA can lead to errors in \Om{} of tens of percent. It has been extensively studied in simulated galaxies \citep{Debattista2003, Guo2019, Zou2019} and in observations \citep{Garma-Oehmichen2020}. In G20, we measured the disc PA by using isophote profiles and by fitting a kinematic model to the H$_\alpha$ velocity map. After discussing the limitations in both methodologies, we concluded that the photometric estimate was more reliable as it produced more measurements that made physical sense. However, our sample was small, and our conclusion based on empirical arguments. 

In this work, we use 3 independent methods for estimating the disc PA with different tracers. (i) The classical photometric approach, using the galaxy isophotes. (ii) Symmetrizing the stellar velocity map using the \code{PAFit} package \citep{Krajnovic2006}. (iii) Modelling the $H_\alpha$ velocity maps using a modified version the package \code{DiskFit} (see Section \ref{sec:rotation_curve_model})  \citep[][Aquino-Ortíz et al. in prep.]{Spekkens2007, Sellwood2010}. We will refer to them as \PAph{}, \PAsym{} and \PAmod{}, respectively.

Each method has biases and systematic errors that cannot be estimated from the mathematical formal errors. For example, to estimate the disc PA from isophotes, we have to assume a thin and circular disc. Strong nearby field stars, light from companion galaxies, disc warps and strong spiral arms can distort the photometric measurements. In some galaxies, the choice of the outer isophotes heavily depends on the author’s criteria. In cases of ambiguity, where the PA profile flattens in multiple sections,  we use isophotes that are more similar with \PAsym{} and \PAmod{}.

The symmetrized velocity fields can be affected by local deviations from axi-symmmetry \citep{Krajnovic2011}. In particular, the non-axisymmetric motions produced by the bar and other structures affect the stellar kinematics \citep{Stark2018}. In some galaxies, the IFU data only covers the central region of the galaxy, ignoring the large-scale movements of the disc.

Modelling the velocity field of the gas has the advantage of using a tracer that is not affected by the asymmetric drift present in old stellar populations. However, the $H_\alpha$ gas can be affected by the AGN feedback, recent galactic encounters, or episodes of strong star formation \citep{Tsatsi2015, Stark2018}. Our kinematic modelling has the additional advantage of disentangling the disc circular movements from the non-axisymmetric movements produced by the bar \citep{Spekkens2007}.

\begin{figure}
    \centering
    \includegraphics[width=\linewidth]{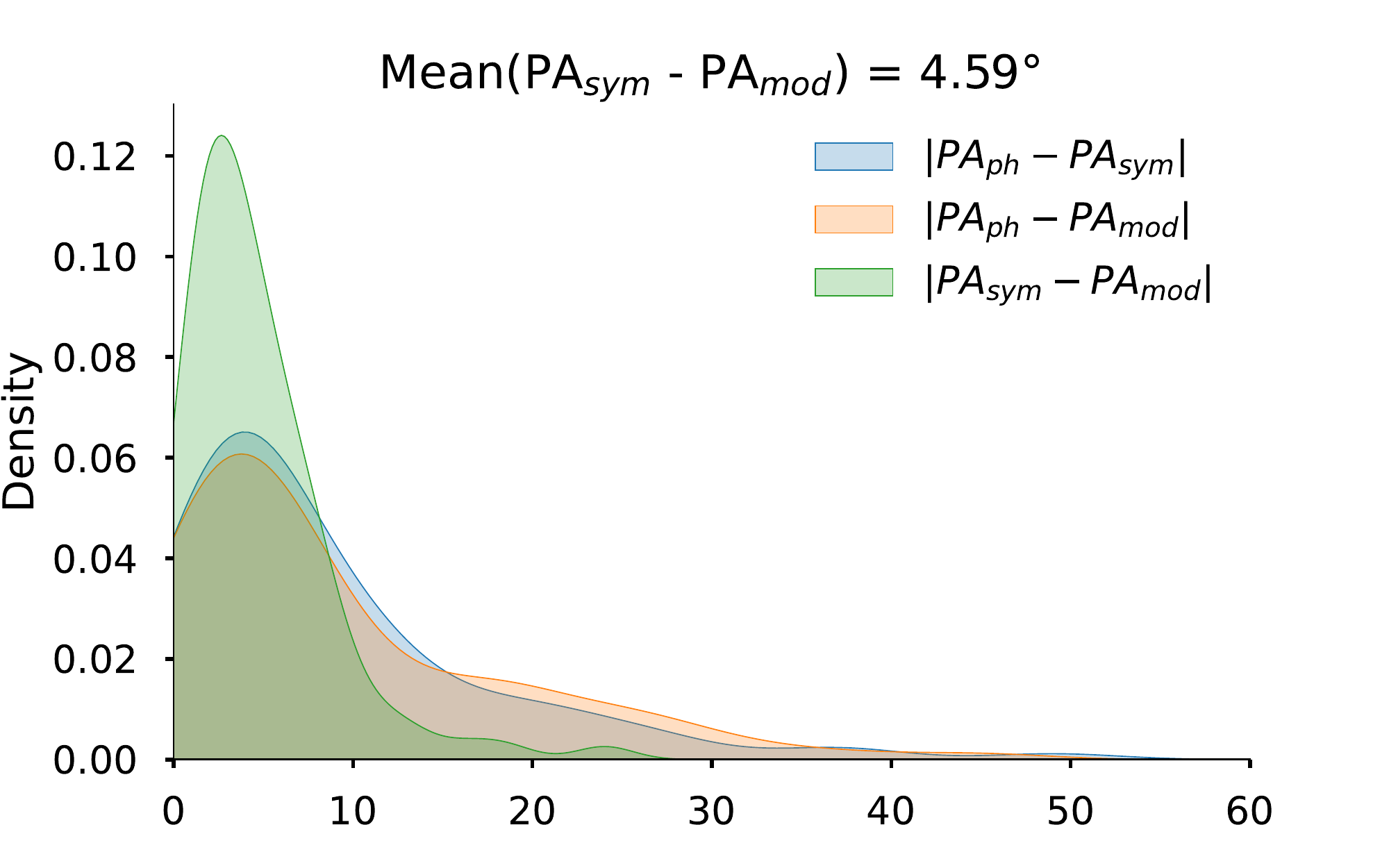}
    \caption{Distribution of the misalignment between three independent PA methods. Although the differences between \PAsym{} and \PAmod{} are smaller, they are significant enough for the TW method.}
    \label{fig:PA_diff}
\end{figure}

In most galaxies of our sample the three measurements are not consistent within their formal uncertainties. In Figure \ref{fig:PA_diff} we show the distribution of the absolute difference between the three methods for all galaxies in our sample. It is not surprising that \PAsym{} and \PAmod{} are the most similar, as both are based on the galaxy kinematics \citep{Barrera-Ballesteros2014}. Nonetheless, the average difference between both is $\sim 4.5 \degree$, which is enough to produce significant differences in \Om{}. The large misalignment between the rotation of stars and gas is possibly provoked by environmental effects like mergers, gas accretion and interaction with nearby galaxies \citep{Khim2020, Khim2021, Lu2021}. (See also \cite{Jin2016}).

In order to weight the 3 PA measurements the following procedure was implemented: 
(i) We assume that the systematic errors are unknown in all our measurements and that their formal errors can be increased by using a penalisation factor. 
(ii) All uncertainties are increased to be at least $1 \degree$. 
(iii) If all measures agree within their uncertainty, we proceed to use a traditional variance weighted mean.
(iv) If they do not agree, we check the linearity condition between \xint{} versus \vint{} at each PA.
(v) PAs that do not follow a linear trend are penalised by increasing their error by a factor of 10. 
(vi) The PAs that do present linearity have their errors increased to the point where they are compatible within their uncertainties. 
(vii) We use the variance weighted mean for our final estimation of the disc PA.

\begin{figure}
    \centering
    \begin{subfigure}{0.48 \textwidth}  
    \includegraphics[width=\linewidth]{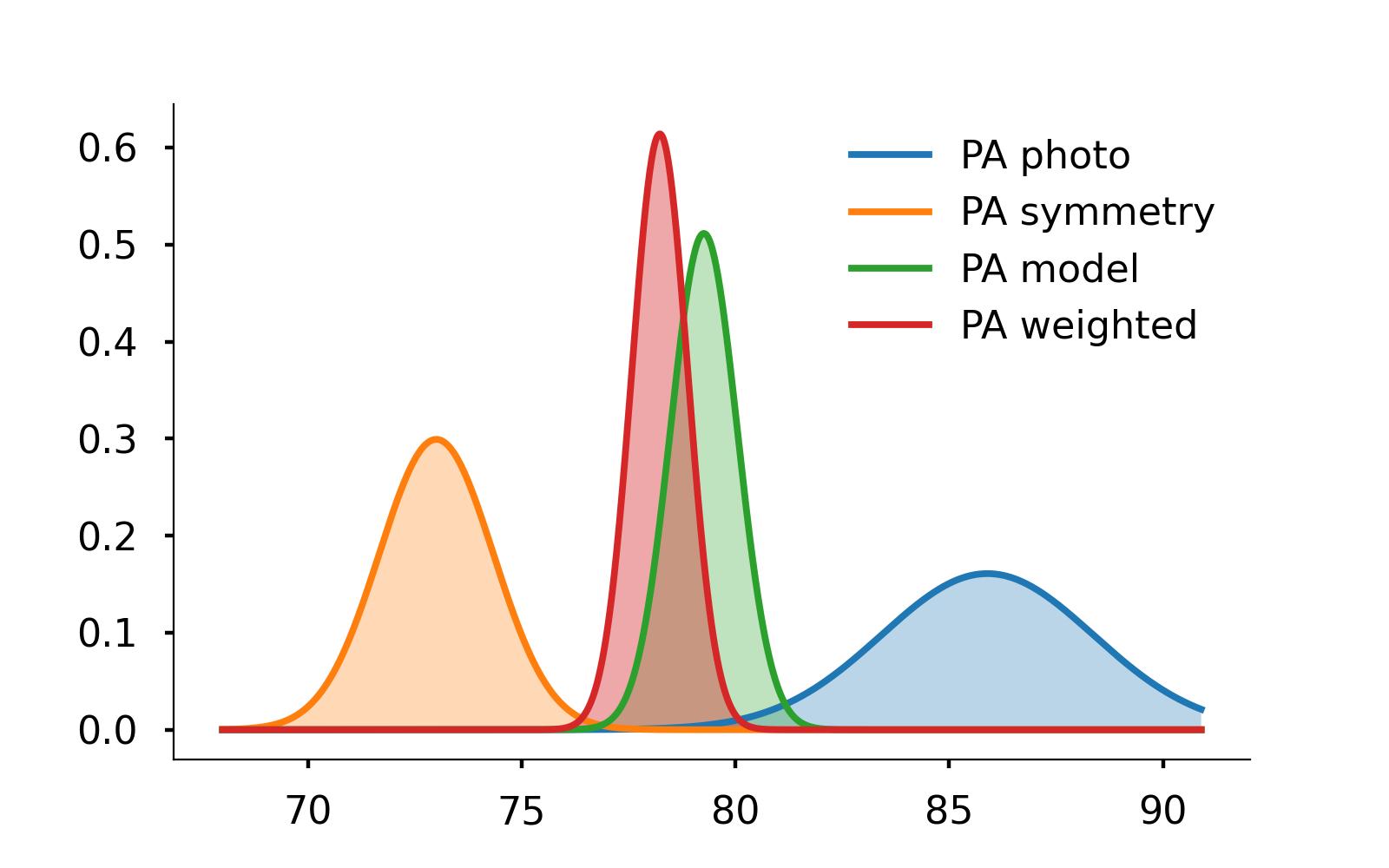}
    \end{subfigure}
    \begin{subfigure}{0.48 \textwidth}  
    \includegraphics[width=\linewidth]{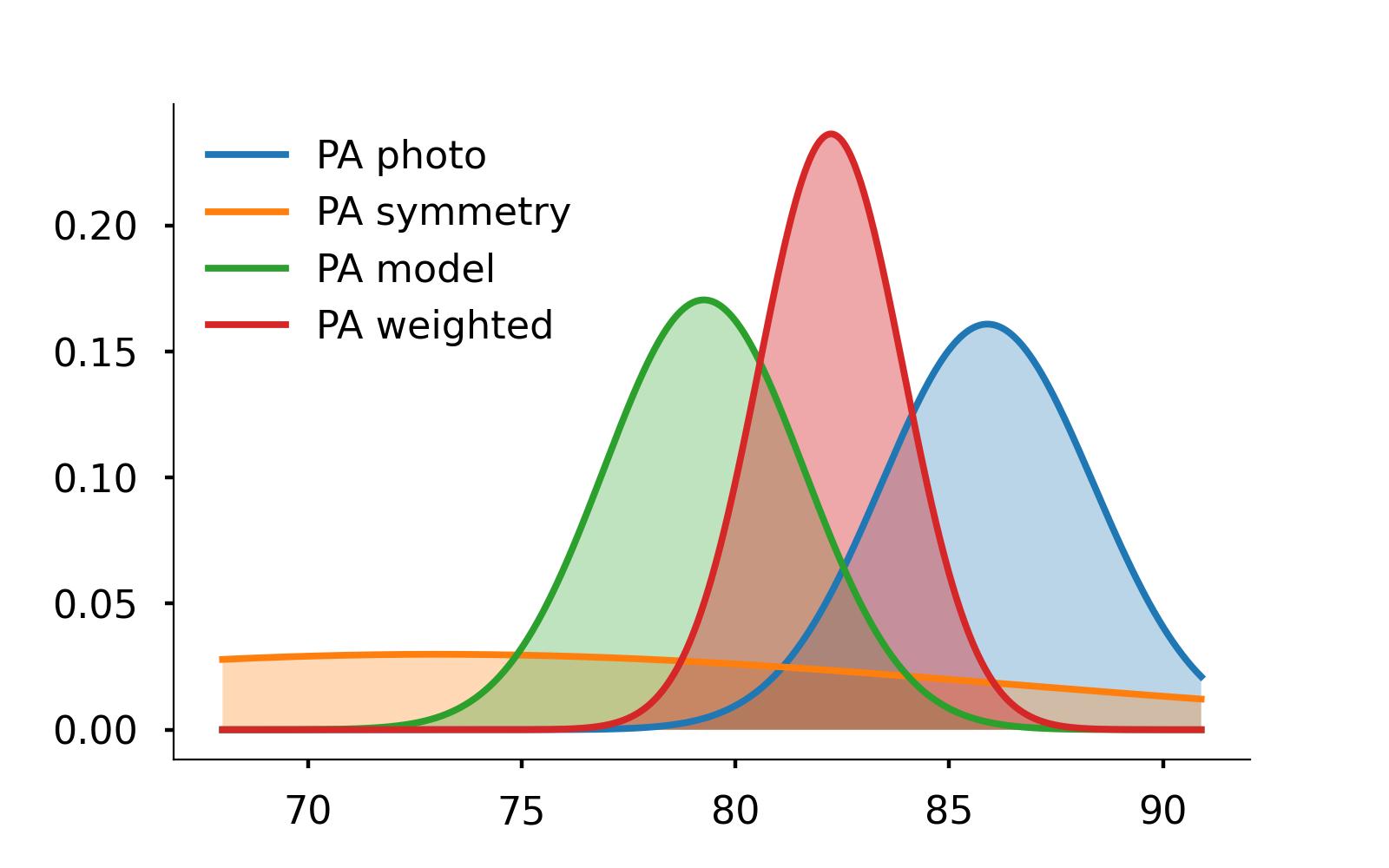}
    \end{subfigure}
    \caption{Weighting the 3 PA measurements. \textit{Top panel:} PA measurements of 8596-12704 before penalising. The three methods do not agree within their 1-sigma uncertainties: \PAsym{} = $73.0 \pm 1.3$,  \PAmod{} = $79.3 \pm 0.8$, \PAph{} = $85.9 \pm 2.48$. \textit{Bottom panel:} PA measurements after penalisation. \PAsym{} = $73.0 \pm 25.0$,  \PAmod{} = $79.3 \pm 2.5$, \PAph{} = $85.9 \pm 2.5$. The TW integrals at \PAsym{} do not follow the lineal trend (see Figure \ref{fig:TW_integrals_wrong_PA}). 
    Thus we choose to penalise it, and use an equal weight between \PAmod{} and \PAph{}.}
    \label{fig:Weighted}
\end{figure}

Figure \ref{fig:Weighted} shows the estimates of PA for our example galaxy before and after penalising (in the top and bottom panels respectively). The differences between PA may be attributed to the strong spirals affecting the estimation of \PAph{}, or to the small field of view of the IFU that only covers the barred and ringed region affecting both \PAsym{} and \PAmod{}. The strong bar of the system, can also be influencing the long scale motions of the stars and the gas.

Since these measurements are not compatible in their uncertainties, we proceed to look at the linearity of the TW diagram. Figure \ref{fig:TW_integrals_wrong_PA} shows the TW diagram measured at \PAsym{} (see Section \ref{sec:pattern-speed}) where the lack of a linear relation between \xint{} and \vint{} is evident. We penalise the measurement by increasing the \PAsym{} error by a factor 10. In comparison, both \PAmod{} and \PAph{} satisfy the linearity requirement so we make them equal weighted. In Figure \ref{fig:TW_integrals} we show the TW diagram measured at the mid-point between both PAs.

In some cases, only one estimate of the disc PA satisfies the linearity condition. In our sample, \PAsym{} tends to produce the better estimations of \xint{} and \vint{} (our example galaxy is an exception), followed closely by \PAmod{}.


\subsection{Bar pattern speed}
\label{sec:pattern-speed}

In G20, we assumed that each error source was independent and could be added in quadrature for our final estimation of \Om{}. This assumption was incorrect, as the behaviour of the TW integrals is non-linear, especially when considering the PA and the slit length. Instead, in this work, we choose to sample all parameters and error sources in the same Monte Carlo procedure.

We modelled the PA as a gaussian distribution with parameters from the weighted measurement of the three PA measurements described in the last section. The galaxy centre was estimated from the centre of mass of the r-band surface-brightness distribution. The mean values and co-variance matrix between $x$ and $y$ coordinates are used to model a 2-D Gaussian distribution. Considering the spatial correlations, the centre distribution looks like an ellipse oriented in the same direction as the bar.

The relative slits length (relative to the maximum slit length that fits the data) was modelled using a half-Gaussian distribution with $\mu = 1$ and $\sigma=0.2$. We use this distribution to ensure most of our measurements use large slits, as the TW method suggests, but also capture some of the systematic error associated with the data coverage. In total, we draw a sample of $\times 10^6$ combinations of these parameters.

The number of slits varies in each galaxy as it depends on the bar coverage, orientation and the IFU size, but on average we use $\sim 15$ slits per galaxy. This results in an average of 15 $\times 10^6$ TW integrals, and 1 $\times 10^6$ measurements of $\Omega_{bar} \times \sin i$ for each galaxy.

\begin{figure}
    \centering
    \includegraphics[width=\linewidth]{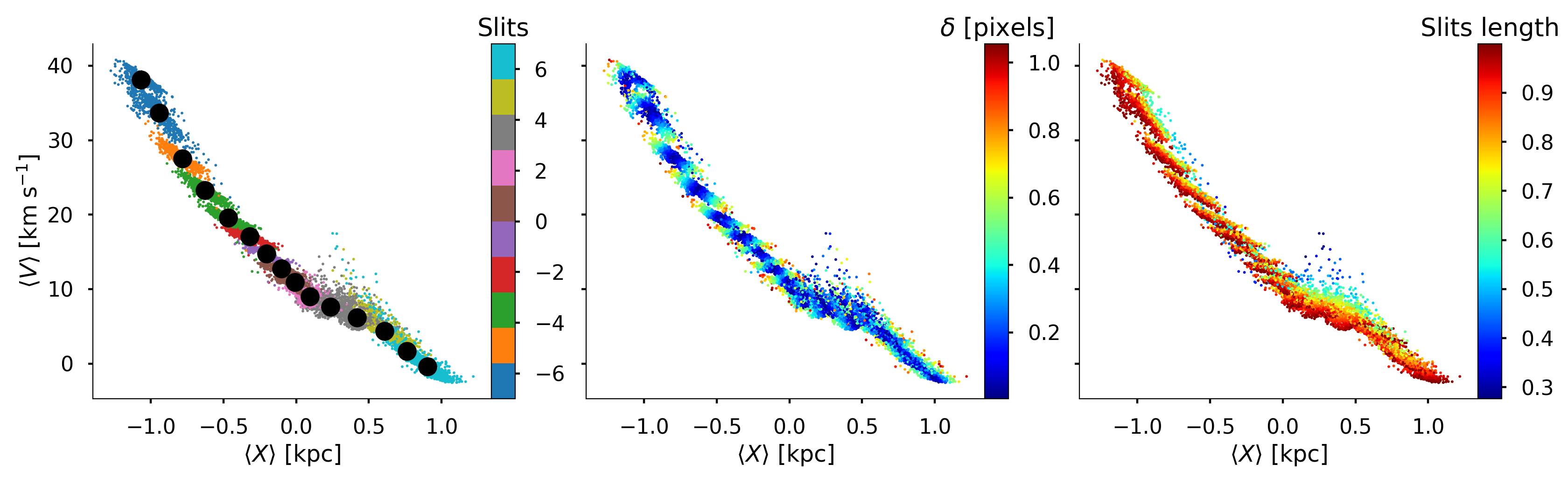} 
    \caption{TW integrals \xint{} and \vint{} measured near the midpoint between \PAph{} and \PAmod{}. Left panel: colour coded by the slit number. The black dots show the median value for each individual slit. Middle panel: coloured by the distance of the random centre to the mean centre of mass. Right panel: coloured by the relative slit length.}
    \label{fig:TW_integrals}
\end{figure}

We can separate the effects of each error source by colouring the TW integrals as a function of the random parameters. In Figure \ref{fig:TW_integrals}, we show the TW diagram of our example galaxy measured near the equal-weighted mean between \PAph{} and \PAmod{}. Each point is a measurement made by an individual slit, with a random centre and random slit length. The slope of the plot is \Om{}$\times \sin i$. The three panels show the same data but are coloured using the slit number (left panel), the distance to the centre (middle panel) and the relative slit length (right panel). For the latter two panels, if the colours are randomly distributed, the measurements do not dependent on the respective error source, while stratified colours show the opposite. In this example, \xint{} and \vint{} have a strong linear relation, and low dispersion due the random parameters.

\begin{figure}
    \centering
    \includegraphics[width=\linewidth]{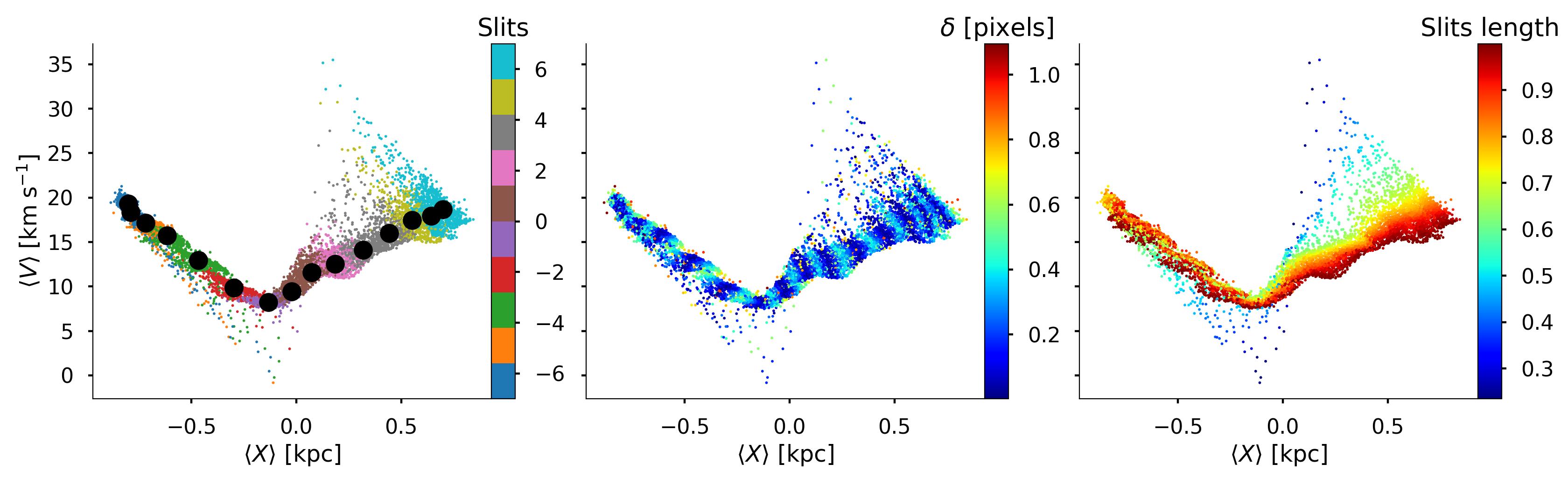} 
    \caption{Same as Figure \ref{fig:TW_integrals}, but measured near \PAsym{}. At this orientation, the measurements highly depend on the slit length.}
    \label{fig:TW_integrals_wrong_PA}
\end{figure}

Figure \ref{fig:TW_integrals_wrong_PA} shows the same plot, but the integrals are computed near \PAsym{}. In this orientation, the TW integrals do not follow the linear trend, and the dispersion of each slit strongly depends on the slits length, as evidenced by the stratification in colours in the third panel. When we observe strange behaviours like this, we penalise the corresponding measurement of PA.

\begin{figure}
    \centering
    \includegraphics[width=\linewidth]{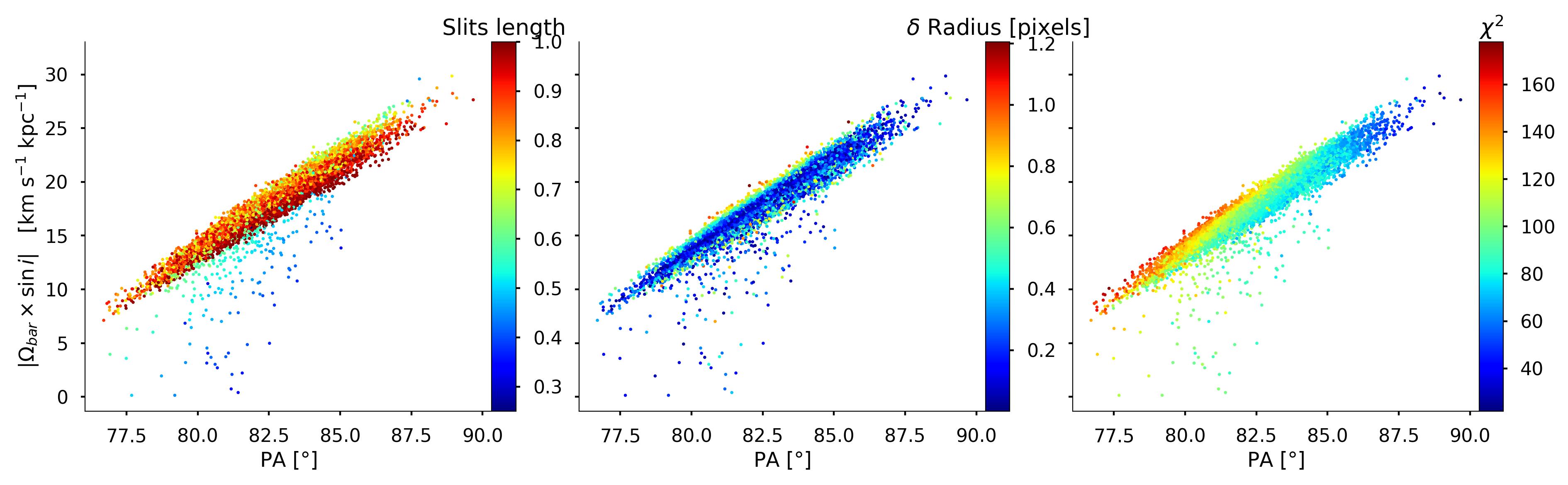}
    \caption{Bar pattern speed versus PA. Left panel: Coloured by the relative slits length. Middle panel: Coloured by distance to the centre. Third panel: Coloured by $\chi^2$ of the fit. Notice how the goodness of fit heavily depends on the PA.}
    \label{fig:om_pa}
\end{figure}

Figure \ref{fig:om_pa} shows the relationship between $\Omega_{bar} \times \sin i$ and the disc PA for our example galaxy. Each point is the slope of a linear fit between \xint{} and \vint{} with a random centre and slit length. Again, we show three panels to disentangle the effects of each error source. The slit length becomes more important at PAs $< 82\degree$ where the shorter slits produce smaller values of \Om{}. In this example, the most important error source is the PA (as usual), as it moves the distribution of \Om{}$\times\sin i$ between 10 to 25 km s$^{-1}$ kpc$^{-1}$. The third panel shows the $\chi^2$ of the linear fit. Notice how the linear fit is getting worse as we move to lower values of PA (closer to PA$_{sym}$).

\begin{figure}
    \centering
    \includegraphics[width=\linewidth]{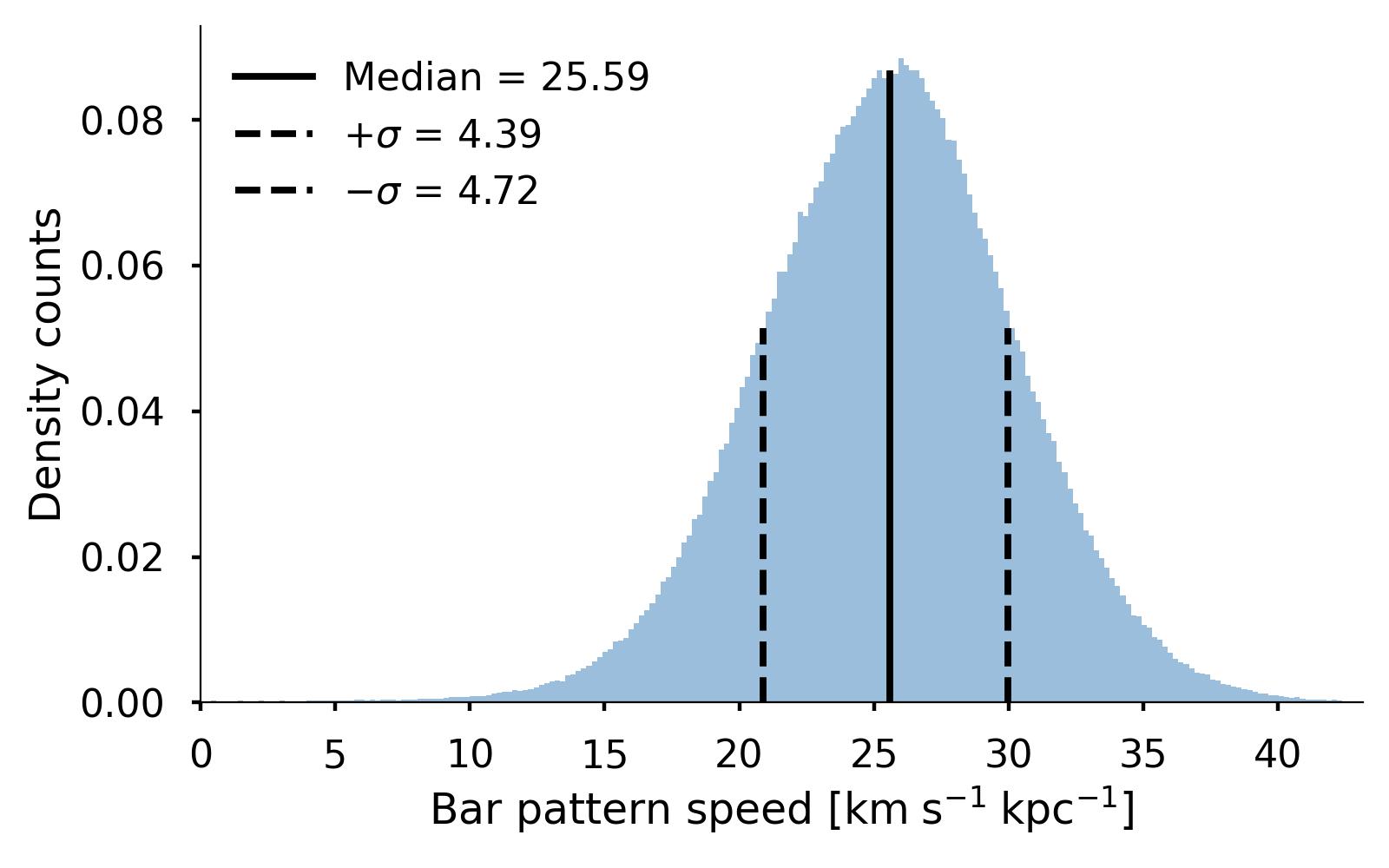}
    \caption{Bar pattern speed distribution of 8596-12704. We used a Monte Carlo procedure where we modelled the disc PA and disc ellipticity as gaussian distributions; the slits length as a half-gaussian distribution; and the galaxy centre as a 2D gaussian distribution.}
    \label{fig:om_dist}
\end{figure}

To get the final distribution for \Om{}, we divide each measurement by a randomly sampled $\sin i$. The inclination is recovered from disc ellipticity using the transformation $i = \arccos (\epsilon)$. We got two measurements of $\epsilon$: (i) from the isophote analysis and (ii) from the H$_\alpha$ kinematic model. As with the disc PA, we expect similar biases to be affecting the disc ellipticity estimations, so we use an equally weighted ellipticity in all our galaxies. The mean and standard deviation of this estimate are used to model $\epsilon$ as a Gaussian distribution. Figure \ref{fig:om_dist} shows the final distribution of \Om{} for our example galaxy.

The inclination uncertainty smooths the distribution of \Om{}. It is important to note that a Gaussian distribution in $\epsilon$ naturally transforms into a right-skewed distribution in $\sin i$. This is relevant in galaxies with high inclination uncertainty  (for example $\Delta i > 6 \degree$), as this results in a right-skewed distribution in \Om{}. We will discuss how the inclination affects the determination of \Rpar{} in Section \ref{sec:where_are_the_ultrafast}.

\subsection{Bar radius}
\label{sec:bar_radius}

There is no consensus on what is the best way to estimate the bar radius. Bars rarely have a well-defined boundary and are often associated with structures like bulges, spirals, rings and ansae. This makes different techniques prone to different biases.

Traditionally, the bar radius is recovered from the isophotal ellipticity and PA profiles. The barred region is characterised by an increasing ellipticity that peaks near the end of the bar and remains constant in PA. The local maximum in ellipticity (hereafter $R_\epsilon$) is a measurement that correlates well with visual estimates \citep{Herrera-Endoqui2015, Diaz2016_Bar_characteristics}. An upper limit can be defined as the radius where the PA has changed $\sim 5 \degree$ from the value at $R_\epsilon$ (hereafter $R_{PA}$) \citep{Aguerri2009}. We can also identify the bar end with a sharp decline in the ellipticity profile. In galaxies where we observe a sharp decline in ellipticity before $R_{PA}$, we use that transition as our upper limit. Figure \ref{fig:isophotes} shows both measurements with a green stripe. In Figure \ref{fig:flux_vel} these estimates are drawn as black ellipses.   

However, the isophote technique is prone to several biases. In some galaxies, the maximum ellipticity is actually produced by open spiral arms that follow the bar. Ansae structures are relatively common in late-type galaxies, and it is not clear if they should be considered part of the bar or the disc \citep{Martinez-Valpuesta2007}. In galaxies with large bars, the choice between logarithmic or linear isophotal steps can alter the measurement of $R_\epsilon$ by a few arcsec. Finally, the bulge-to-total ratio (B/T) can affect the estimation in galaxies with weak exponential bars \citep{Lee2020}. 

Other techniques to measure the bar radius include the ratio of the bar to inter-bar intensity, via Fourier decomposition of the light \citep{Aguerri2000}, a 2D multi-component surface brightness decomposition (fitting the light distribution of the disc + bulge + bar) \citep{Laurikainen2005, Gadotti2011, Salo2015} and the maximum bar torque radius (based on the transverse-to-radial force ratio $Q_T$) \citep{Sanders1980, Combes1981, Diaz2016_Bar_characteristics, Lee2020}.

In G20 we modelled the bar radius using an uniform distribution with $R_\epsilon$ and $R_{PA}$ as the lower and upper limits. In this work we choose to model the bar radius as a log-normal distribution, using $R_\epsilon$ as the mean and $R_{PA}$ as a 2-sigma upper limit. This is:

\begin{equation}
    \mu = \log \left( R_\epsilon \right)
\end{equation}

\begin{equation}
    \sigma = \log \left( \sqrt{ \frac{R_{PA}}{R_\epsilon}} \right)
\end{equation}

This distribution let us explore a smaller bar radius, as suggested by \cite{Hilmi2020} and \cite{Cuomo2021}. The logarithmic distribution also prevents sampling negative bar radius and, when the $\sigma / \mu$ ratio is small, the distribution resembles a Gaussian distribution.    

We deproject the bar using the analytical procedure described in \cite{Gadotti2007}. The method works well in galaxies with moderate inclination angles ($<60 \degree$) \citep{Zou2014}, which is the case for most galaxies in our sample . The method requires the relative orientation of the bar with the disc, as well as the disc inclination angle.

\subsection{Corotation radius and the H$_\alpha$ rotation curve model}
\label{sec:rotation_curve_model}

\begin{figure}
    \centering
    \begin{subfigure}{0.48 \textwidth} 
    \includegraphics[width=\linewidth]{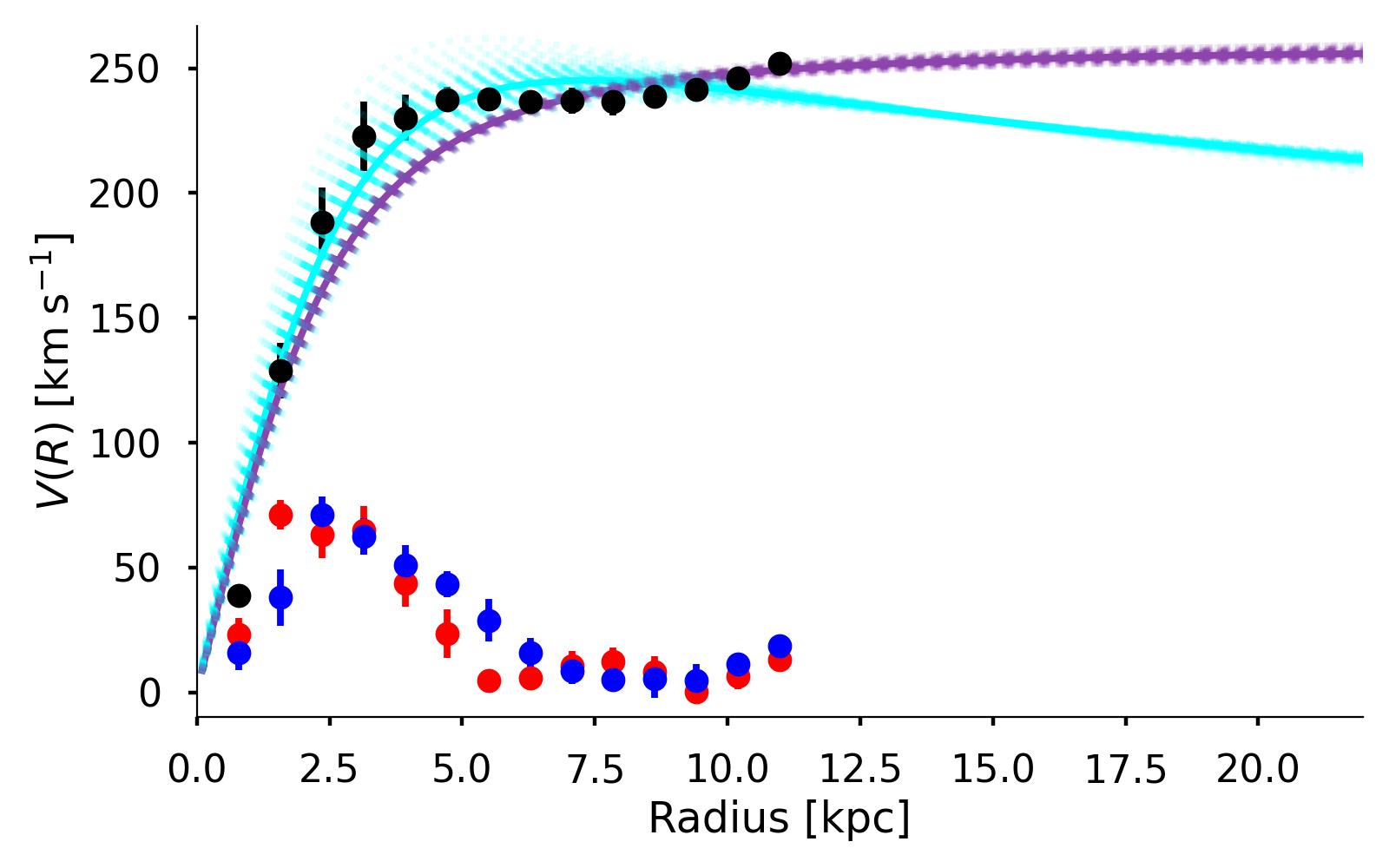}
    \end{subfigure}
    \begin{subfigure}{0.48 \textwidth} 
    \includegraphics[width=\linewidth]{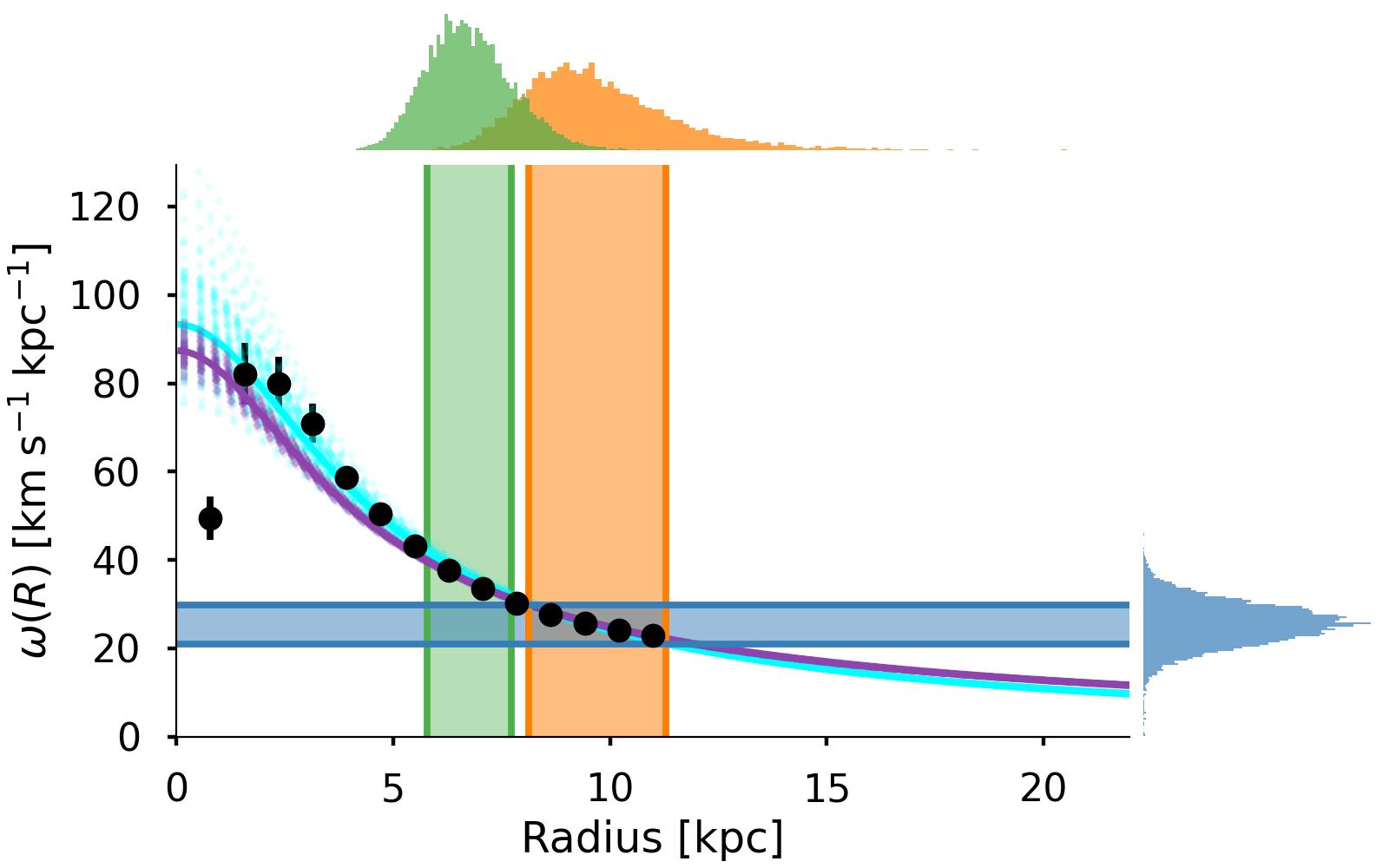}
    \end{subfigure}
    \caption{Rotation model of 8596-12704. Top panel: Rotation curve. The black dots show the mean circular velocity of the \code{Diskfit} model. The blue and red dots show the fitted radial and tangential non-circular motions. The cyan and purple curve are the best fit of equation \ref{eq:Bertola} using $gamma$ as a free parameter or fixing it to $gamma=1$ respectively. Bottom panel: Angular velocity curve. The bar pattern speed distribution is shown in blue. The bar radius distribution is shown in green. The resulting corotation distribution is shown in orange. The projections are limited by the 1-sigma errors.}
    \label{fig:rotation_curve}
\end{figure}

To get a reasonable measurement of corotation, we require a good estimate of the rotation curve. Several works assume a quick flattening, and estimate \CR{} as the ratio $V_{c}^{flat}/\Omega_{bar}$ where $V_{c}^{flat}$ is the flat circular velocity of the disc \citep{Aguerri2015, Cuomo2019, Guo2019}. Recovering the details in the shape of the rotation curve requires careful modelling. Massive (brighter) galaxies typically display slowly declining rotation curves, while galaxies with low masses have monotonically rising rotation curves \citep{Persic1996, Kalinova2017}. Moreover, bumps and wiggles are common features of rotation curves, that result from the perturbed kinematics of the galactic disc components. The amplitude of these wiggles varies depending on the galaxy and perturber (bar, spiral arms, satellites), and takes a proper model to be correctly represented. In G20, we got a relative difference of 21\% in the \CR{} estimation between modelling the rotation curve and assuming a flat disc.

Recovering the rotation curve from 2D velocity maps is straightforward when the tracers have small departures from circular orbits \citep{Begeman1987}. However, stellar bars and other non-axisymmetric features can drive large non-circular motions that complicate the determination of the rotation curve \citep{Valenzuela2007}. 

In this work, we use the code \code{Diskfit} that describes the velocity field with the so-called bi-symmetric model \citep{Spekkens2007, Sellwood2010}. This kinematic model assumes that most perturbations can be described by an $m=2$ order harmonic decomposition. The code yields the mean circular velocities and amplitudes of non-circular streaming velocities in the radial and tangential directions.

The code \code{Diskfit} uses a Levenberg \- Marquardt (LM) algorithm as a minimisation technique. However, the parameter space may have several local minimum values, where the LM routine can easily get trapped. To overcome this issue, we used a modified version that implements a Markov Chain Monte Carlo method (Aquino-Ortíz et al., in prep). The input parameters are optimised using the Metropolis-Hastings algorithm \citep[e.g.][]{Puglielli10}. This version has been used in recent kinematic studies \citep[e.g.][]{Aquino-Ortiz2018, Aquino-Ortiz2020}.

In most galaxies of our sample, the kinematic model is not enough to estimate \CR{}. Because of the data coverage, the rotation curve only covers the central region of the galaxy, and additional modelling is required to extrapolate the outer parts. We use a 3 parameter model \citep{Bertola1991}:

\begin{equation}
V(r) = \frac{V_c r}{(r^2 + k^2)^{\gamma / 2}}
\label{eq:Bertola}
\end{equation}

where $V_c$ controls the amplitude, $k$ the sharpness and $\gamma$ the slope at large radius. When $\gamma >1$ the model corresponds to a declining curve and $\gamma < 1 $ to a rising curve. It is important to note that there is a large degeneracy between $V_c$ and $\gamma$, and in some galaxies, a better fit can be achieved by fixing $\gamma=1$ (a 2 parameter flat rotation model).

Some galaxies require no extrapolation, as \CR{} can be determined within the kinematic model. In those cases, we used a spline fit that recovers small details like wiggles that cannot be modelled with the parametric functions. 

For each measurement of \Om{} we generate a random rotation curve with the parameters of the best fit. The top panel of Figure \ref{fig:rotation_curve} shows the rotation curve of the example galaxy. The black dots correspond to the mean circular velocity from the \code{Diskfit} model. Blue and red dots at the bottom are the m=2 radial and tangential motions, respectively. The model stops at $\sim$ 11 kpc due to the data coverage. The cyan and purple curves show random iterations of equation \ref{eq:Bertola} where the difference is using $\gamma$ as a free parameter or fixing it to 1. Here, the 3 parameter model fits a declining rotation curve, while the flat model remains at $V_c = 250$ km s$^{-1}$. At large radii, the difference between the two curves becomes more important. 

The bottom panel of Figure \ref{fig:rotation_curve} shows the angular velocity curve. The blue marginalised distribution in the right is the bar pattern speed we obtained in Figure \ref{fig:om_dist}. The green distribution is the bar radius we modelled in Section \ref{sec:bar_radius}. The corotation radius distribution is shown in orange, and was obtained form the intersections of \Om{} with the 3 parameter random angular velocity curves.

\subsection{Rotation rate \Rpar{}}
\label{sec:ParameterR}



To estimate \Rpar{} we divide each element of \CR{} with a random bar radius modelled with a log-normal distribution. The shape of \Rpar{} distribution is usually skewed to the right. It is common that galaxies with large uncertainties (in PA, inclination or bar radius) will show a large tail in \Rpar{} in the slow bar regime.

\begin{figure}
    \centering
    \includegraphics[width=\linewidth]{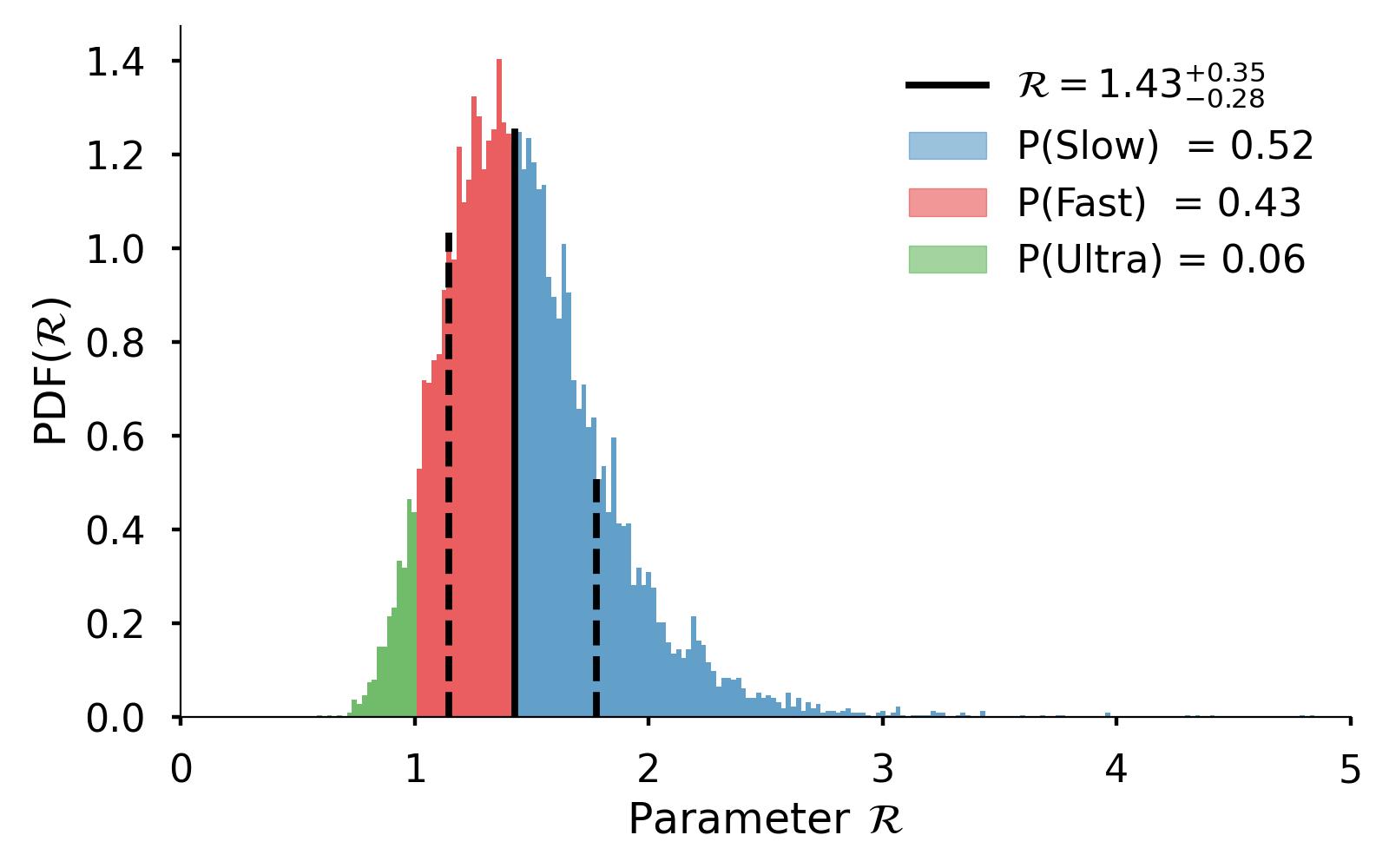}
    \caption{Rotation rate \Rpar{} of 8596-12704. The distribution is coloured using the kinematic classifications of the bar. The area under the curve of each classification is shown in the top right.}
    \label{fig:ParR_example}
\end{figure}

In Figure \ref{fig:ParR_example} we show the resulting probability distribution of the example galaxy.We estimate the probability of bar being slow, fast or ultra-fast by using the area of the probability distribution. In this particular case, the median of the distribution (\Rpar{}$ = 1.43 $) and the most probable classification (Slow = 0.52) both coincide with a slow rotating bar.

\section{Results}
\label{sec:results}

In this section we present the statistical results of our complete sample. We include measurements of global properties from the \code{Pipe3D} package \citep{Sanchez2018} including stellar mass, molecular gas mass, stellar spin parameter and stellar surface density. 



\subsection{Statistics of the full sample}
\label{sec:statistics}

\begin{figure}
    \centering
    \begin{subfigure}{ \linewidth}  
        \includegraphics[width=\linewidth]{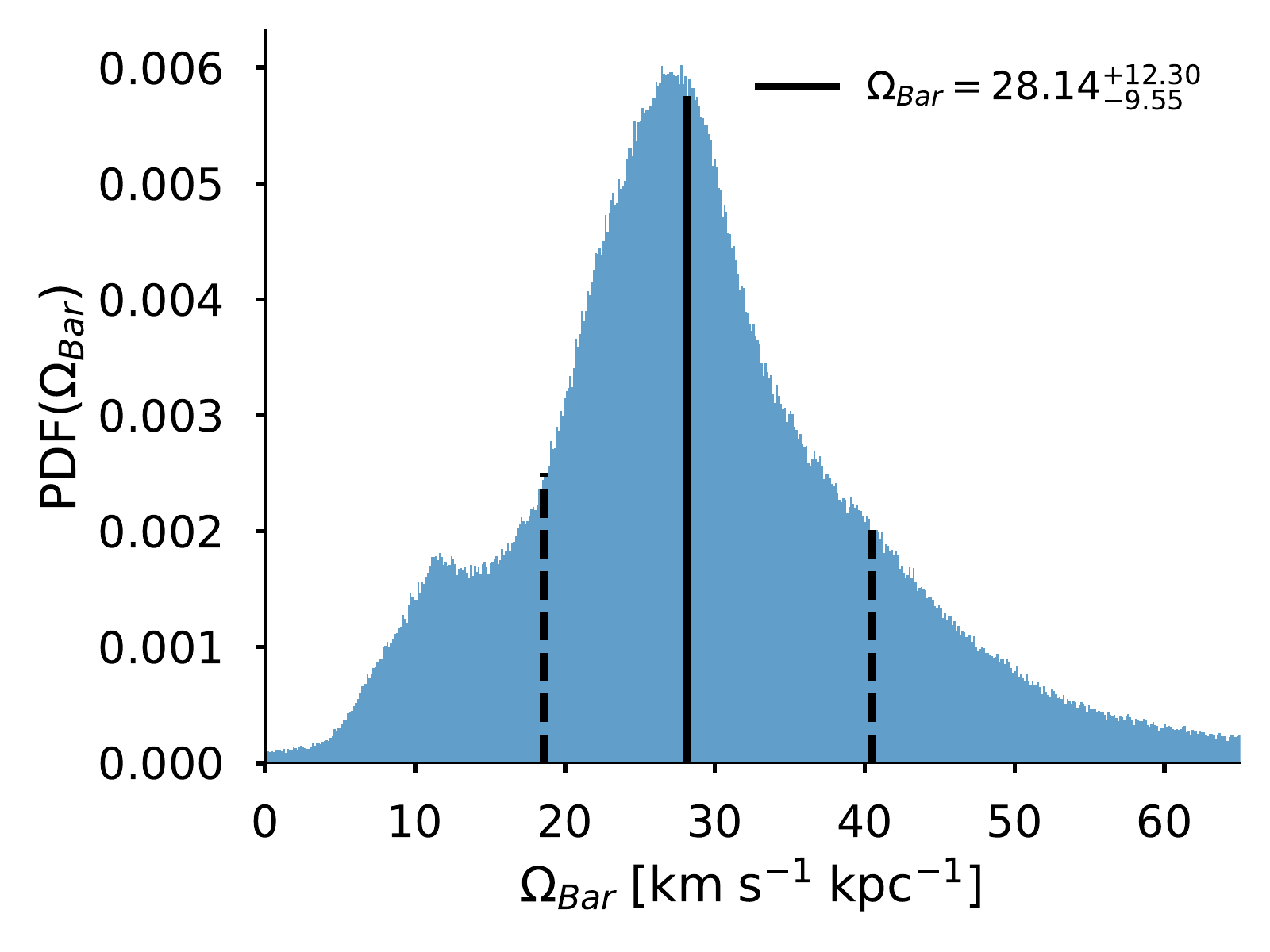}
    \end{subfigure}
    \begin{subfigure}{ \linewidth}  
        \includegraphics[width=\linewidth]{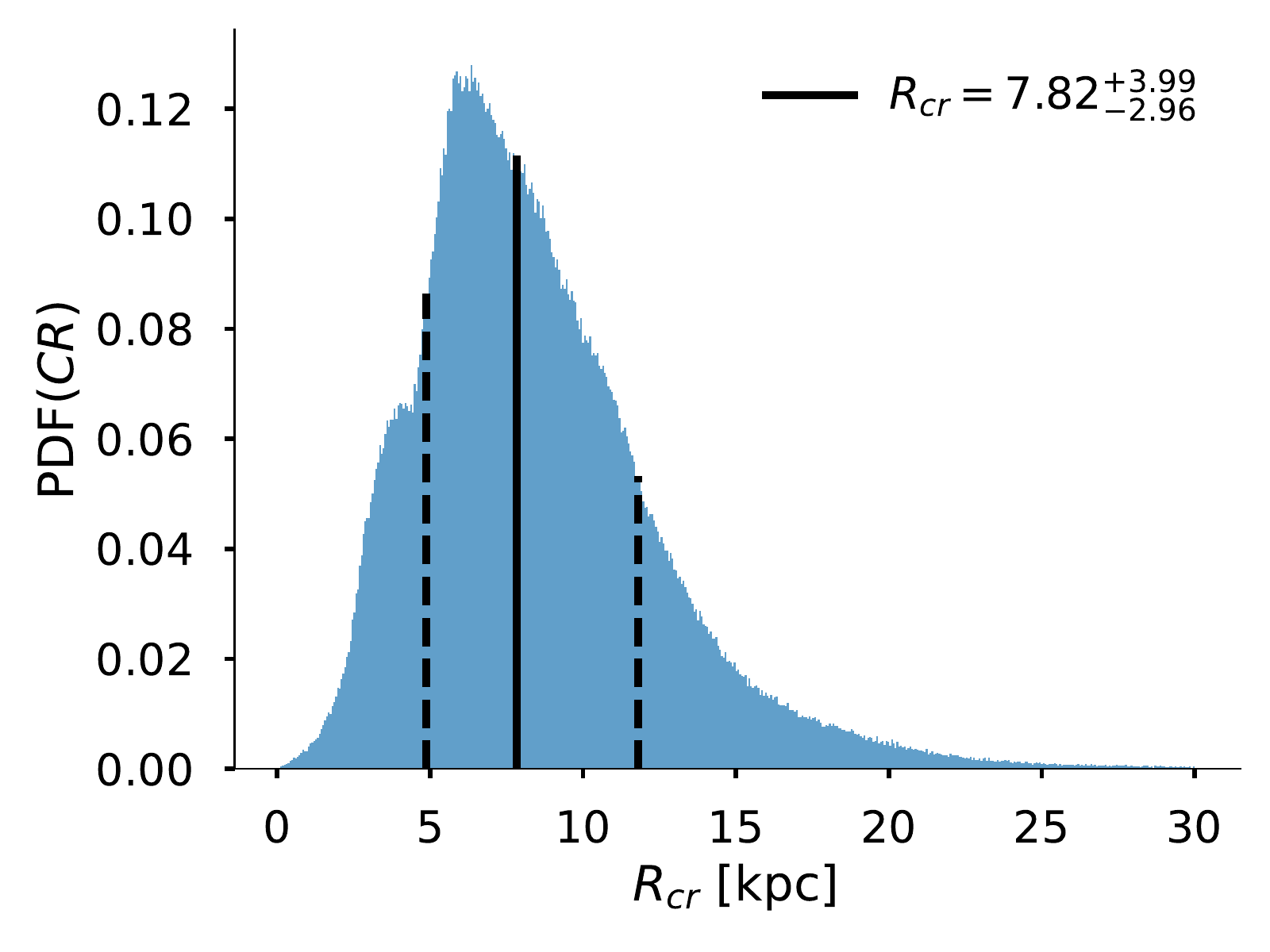}
    \end{subfigure}
    \begin{subfigure}{ \linewidth}  
        \includegraphics[width=\linewidth]{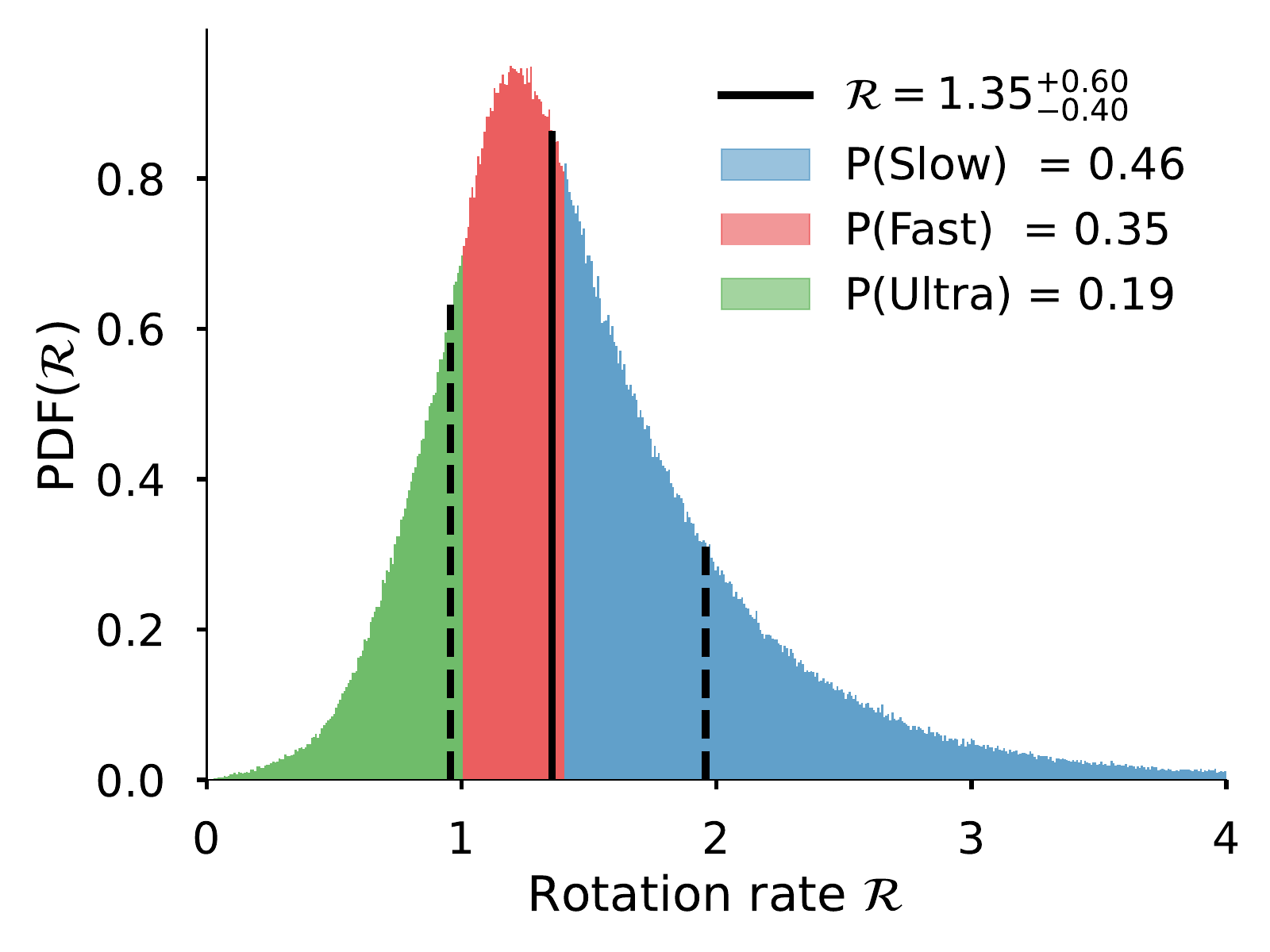}
    \end{subfigure}
    \caption{Distributions of the bar properties for the complete sample of galaxies. \textit{Top panel}: Bar pattern speed. \textit{Middle panel}: Corotation radius. \textit{Bottom panel}: Rotation rate \Rpar{}.}
    \label{fig:om_dist_complete}
\end{figure}

Figure \ref{fig:om_dist_complete} shows the distribution of \Om{}, \CR{} and \Rpar{} after adding the results of the complete sample. The median relative error is 20\%, 22\% and 26\%, respectively. This is slightly smaller than other TW based measurements. For reference, Table \ref{tab:median_errors} shows the median relative errors from \cite{Aguerri2015}, \cite{Guo2019} and \cite{Garma-Oehmichen2020}. Nonetheless, our treatment is not free of biases, and we comment on further improvements for future measurements in Section \ref{sec:improving_measures}. 

\begin{table}
    \centering
    \begin{tabular}{cccc}
    \hline \hline
      Work & \Om{} & \CR{} & \Rpar{} \\
       (1) & (2) & (3) & (4) \\
      \hline
    Aguerri et al. 2015 & 35\% & 43\% & 39\%   \\
    Guo et al. 2019 &  24\%  & 28\% & 37\% \\
    Garma et al. 2020 & 18\% & 30\% & 30\% \\
    This work & 20\% & 22\% & 26\% \\
     \hline
    \end{tabular}
    
    \caption{Median relative errors in TW based measurements. For all works we used the light-weighted results at the photometric PA orientation. Col. (1) Work. Col. (2) Relative error in the bar pattern speed. Col. (3) Relative error in corotation radius. Col. (4) Relative error in rotation rate.}
    \label{tab:median_errors}
\end{table}

\begin{figure}
    \centering
    \includegraphics[width=\linewidth]{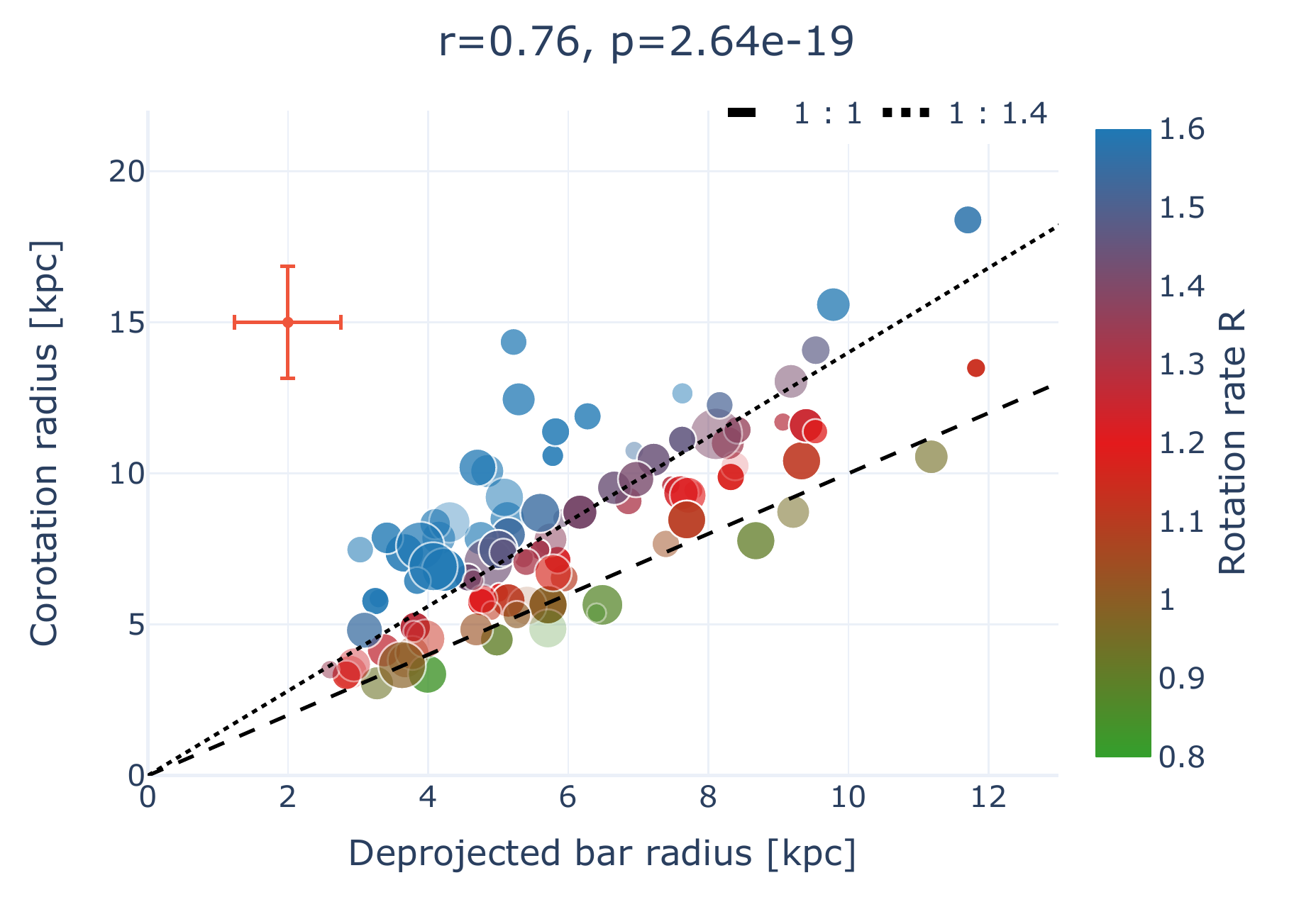}
    \caption{Deprojected bar radius versus corotation radius. The dots size and opacity are used to illustrate the uncertainty in the bar radius and corotation, respectively. We show the average error bar in red. The ratios 1:1 and 1:1.4 used to separate ultra-fast, fast and slow bars are shown with the segmented and dotted lines respectively.}
    \label{fig:bar_med_cr_med}
\end{figure}

Most of our sample is near the border between fast and slow bars. Using the most probable classification from the distribution of \Rpar{}, our sample is composed of 52 slow, 26 fast and 19 ultra-fast bars. Figure \ref{fig:bar_med_cr_med} shows the deprojected bar radius versus corotation radius of all our sample, coloured by \Rpar{}.

To improve the visualisation of the following figures, we opt to represent the $x$ axis uncertainty with the dots size and the $y$ axis uncertainty with the opacity. Thus, the more diffuse-transparent (compact-solid) dots correspond to our most uncertain (certain) measurements. For reference, we show in red the average error bar. We also include in the title the Spearman correlation coefficient and the corresponding $p$-value. All measurements can be found in the Appendix table \ref{tab:master} and the public repository (See Data Availability statement).

\subsection{Correlations}
\label{sec:correlations}

\begin{figure}
    \centering
    \includegraphics[width = \linewidth]{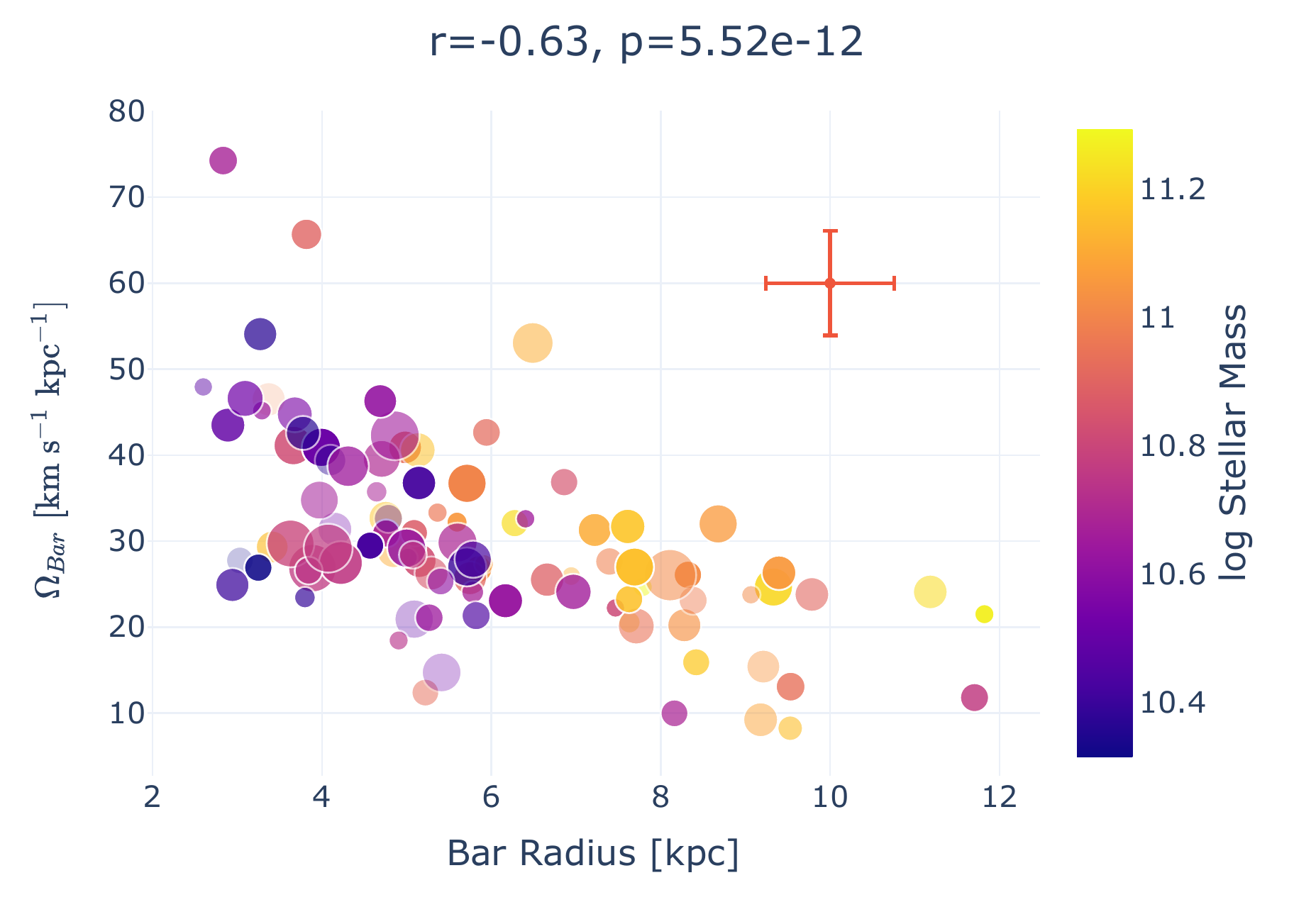}
    \caption{Deprojected bar radius versus bar pattern speed coloured by the logarithmic stellar mass.}
    \label{fig:bar_med_om_med}
\end{figure}

Many works have tried to found correlations between the bar pattern speed and other bar parameters. The strongest of these relations usually occurs between \Om{} and the bar length $R_{bar}$ (longer bars rotate with lower pattern speeds). For example in G20 we found a Spearman correlation coefficient of \Spear{-0.53} with a sample of 18 galaxies. Later, using a sample of 77 galaxies with direct measurements of \Om{} from different pieces of literature, \cite{Cuomo2020} found a stronger relation with \Spear{-0.64}. In this work, we found a similar correlation (\Spear{-0.63}) as shown in Figure \ref{fig:bar_med_om_med}. The relation is best seen when coloured by the stellar mass, as it also reveals the well known bar length - mass relation (\Spear{0.62} in our sample). Interestingly, the relation between \Om{} and the stellar mass is not nearly as strong (we get a weak relation with \Spear{-0.24}).

\begin{figure}
    \centering
    \includegraphics[width = \linewidth]{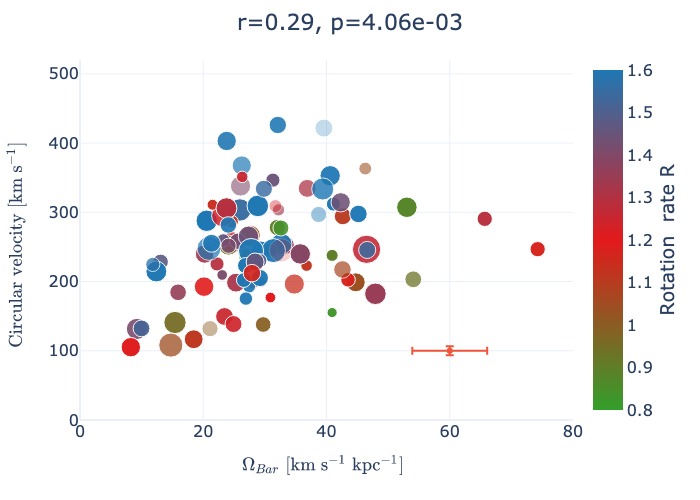}
    \caption{Bar pattern speed versus disc flat circular velocity coloured by their rotation rate \Rpar{}. Discs with large values of circular velocity host bars with high bar pattern speed, but are slower in \Rpar{}}
    \label{fig:om_med_vc_flat}
\end{figure}

\begin{figure}
    \centering
    \includegraphics[width = \linewidth]{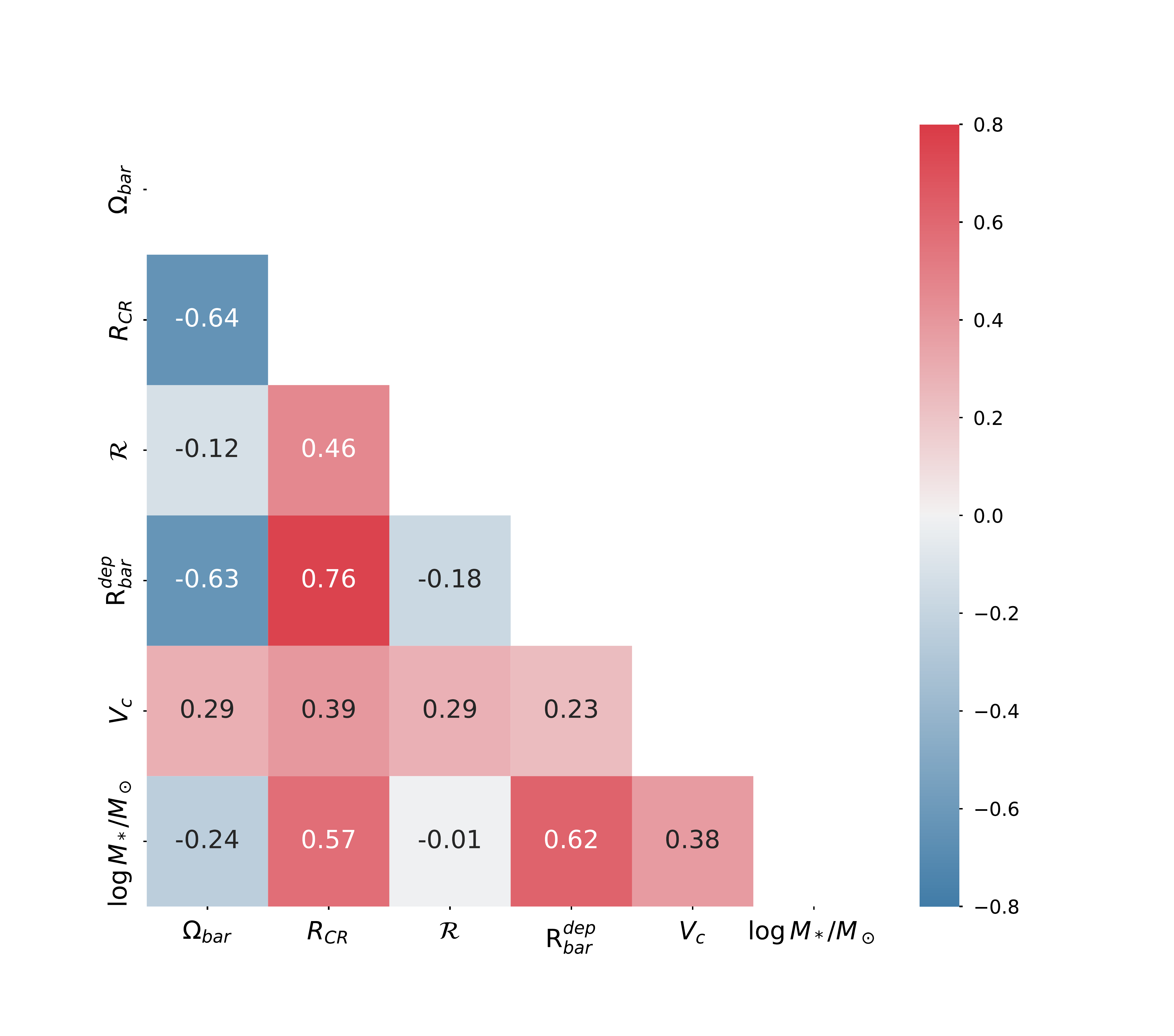}
    \caption{Spearman correlation matrix of quantities shown in Figures \ref{fig:bar_med_om_med} and \ref{fig:om_med_vc_flat}}
    \label{fig:correlation}
\end{figure}

There is another global parameter that is positively correlated with \Om{}: the disc flat circular velocity (\Spear{0.29)}. In Figure \ref{fig:om_med_vc_flat} we show this relation coloured by \Rpar{}, which is also positively correlated with $V_c^{flat}$ (\Spear{0.29}). These two correlations suggest that discs with large values of circular velocity will host bars with high bar pattern speed, but will be slower in \Rpar{}. The correlation matrix between all quoted quantities is shown in Figure \ref{fig:correlation}.



In G20 we reported two strong correlations between the molecular gas fraction $f_g = M_{gas}/(M_* + M_{gas})$ with \Rpar{} (\Spear{0.54}) and with \Om{} (\Spear{-0.52)}. In this work, we found that the relation with the rotation rate is still present, but with more scatter (\Spear{0.23}), and that the relation with \Om{} disappeared (\Spear{0.07}). It is possible that our previous work was affected by low number statistics, or was biased by some outliers. Although weak, the relation with \Rpar{} does suggest that the gas role in the bar evolution is important and should be explored in more detail.


We also looked for correlations with other global galactic properties derived from the \code{Pipe3D} analysis \citep{Sanchez2018}. Most of these relations are weak, and have a lot of scatter, but we mention two that could be of interest: the Sérsic index has a weak anti-correlation with \Om{} (\Spear{-0.20}), and \Rpar{} (\Spear{-0.17}). Similarly, the rotation velocity-to-velocity dispersion ratio within 1 effective radius $v/\sigma$ has a weak correlation with \Om{} (\Spear{0.24}), and \Rpar{} (\Spear{0.12}). These two pairs of correlations suggest that more concentrated mass distributions (or more pressure supported systems) produce bars that rotate at lower bar pattern speed, but faster in \Rpar{}. Nonetheless, both $n$ and $v/\sigma$ are not corrected for the presence of the bar. A more realistic study would require such corrections \citep{Gadotti2008, Weinzirl2009, Graham2009}.


\subsection{Where are the ultra-fast bars?}
\label{sec:where_are_the_ultrafast}


In recent years, the abundance of fast-rotating bars has attracted attention from the cosmological simulations community. The angular momentum exchange in simulations produces high rates of bar slowdown. \cite{Roshan2021} found mean values for \Rpar{} range between 2.5 to 3 in the cosmological simulations IllustrinsTNG and EAGLE. \cite{Fragkoudi2021} showed that galaxies more baryon-dominated can remain fast down to z=0. Moreover, \cite{Frankel2022} showed that simulated bars are shorter rather than slower compared to their observed counterparts. Alternatively, \cite{Roshan2021_mond} showed that bars remain fast in modified gravity theories.

Our sample has a strong component of ultra-fast bars (using the probability distributions, they account for $\sim 20\%$ of our sample). Before attempting to explain their physical origin, we explore other possible systematic errors. As mentioned in Section \ref{sec:TW}, the TW method is susceptible to fictitious signals in galaxies with low and high inclinations. Face-on galaxies have poor kinematic information, while edge-on galaxies do not have positional information. Also, the symmetry of the TW-integrals requires the bar not to be oriented towards the minor or major axis of the disc. 

\begin{figure}
    \centering
    \includegraphics[width = \linewidth]{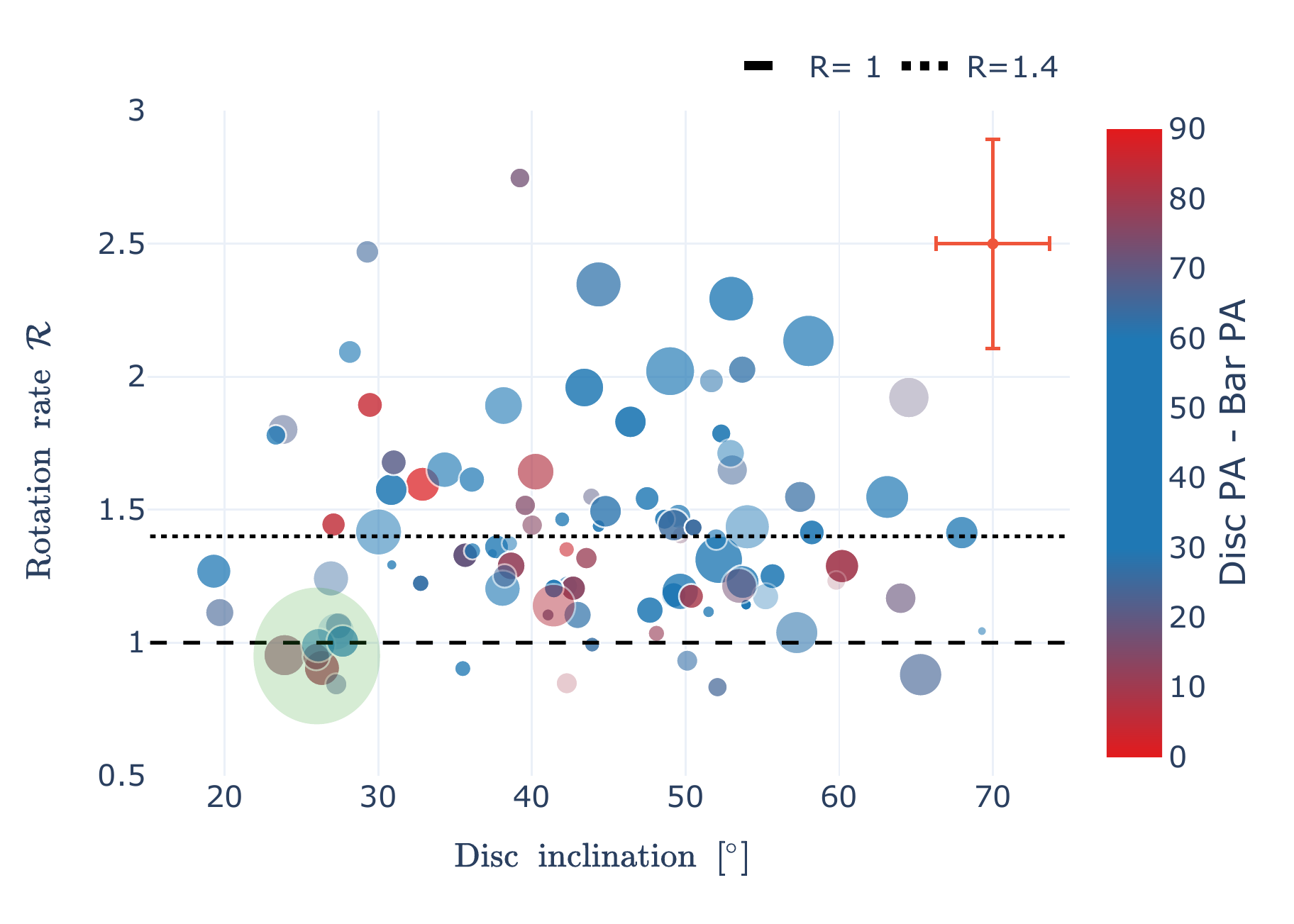}
    \caption{Rotation rate \Rpar{} versus disc photometric inclination coloured by the relative orientation between the disc and the bar. A cluster of ultra-fast bars is found in the low-inclination region, which we highlight with a green circle.}
    \label{fig:inc_R_med}
\end{figure}

Figure \ref{fig:inc_R_med} shows the rotation rate versus the disc photometric inclination. The figure is coloured by the relative orientation of the disc and the bar, highlighting in red galaxies where the relative orientation could be problematic. In particular, we noticed a cluster of ultra-fast bars in the low inclination region (green circle).

We create a \quotes{quality} sub-sample by introducing the following constrains: inclination within $30 \degree < i < 60 \degree$ and relative orientation $20 \degree < | PA_{disc} - PA_{bar} | < 70 \degree$. This quality sub-sample consists of 55 galaxies, this time with only 6 ultra-fast bars ($\sim 10\%$ of the sample). In other words, 14 of our ultra-fast galaxies were probably the result of systematic errors related to the geometric limitations of the TW-method.

Notably, this quality sub-sample also has stronger correlations compared to the complete sample. In table \ref{tab:mw_sample} we compare the correlations discussed in Section \ref{sec:correlations} between both samples.

\begin{table}
    \centering
    \begin{tabular}{ccc}
    \hline \hline
      Relation & Complete Sample & Quality sub-sample \\
       (1) & (2) & (3) \\
      \hline
    \Om{} - $\log M_*/M_\odot$  & -0.23 & -0.46 \\
    \Om{} - $R_{bar}^{Dep}$ & -0.63 & -0.71  \\
    $\log M_*/M_\odot$ - $R_{bar}^{Dep}$ & 0.58 & 0.69  \\
    \Om{} - $R_{cr}$  & -0.64 & -0.75 \\
    \Om{} - $V_c$  & 0.29 & 0.14 \\
    \Rpar{} - $V_c$  & 0.29 & 0.25 \\
    \Om{} - $n$ & -0.20 & -0.31 \\
    \Rpar{} - $n$ & -0.17 & -0.11 \\
    \Om{} - $v/\sigma$ & 0.24 & 0.41 \\
    \Rpar{} - $v/\sigma$ & 0.12 & 0.06 \\
     \hline
    \end{tabular}
    
    \caption{Spearman correlation coefficients in the complete sample and the quality cut sub-sample. Col. (1) Pair of variables. Col. (2) Correlation coefficients in the complete sample Col, (3) Correlation coefficients in the quality sub-sample.}
    \label{tab:mw_sample}
\end{table}

It is possible that some galaxies in our sample have more than one pattern speed (for example from the spiral arms or a nested bar). Although we choose the number of slits to cover only the bar region, we cannot disentangle the effects that other non-axisymmetric structures could have in the estimation of \Rpar{}. Some slow bars in our sample could be affected by the spiral pattern, however it is hard to quantify.


\subsection{Milky Way like galaxies}
\label{sec:mw_like}

Because of the restrictive nature of the TW method, the selection criteria in our sample was based only upon the morphological type and stellar mass. However, our sample contains galaxies with a great variety of bars sizes, morphological features (rings, spirals), mass distribution (with disc circular velocities that range from 150 to 400 km s$^{-1}$) and galactic environments.

To better reflect the parameters of the MW we build a sub-sample making an additional cut in the bar length ($R_{bar} < 6$ kpc) and the disc circular velocities ($190 < V_c < 290$ km s$^{-1}$). We chose these parameters as they are correlated with \Om{}. We also include the cuts in inclination and relative orientation of the disc-bar discussed in the previous section to improve the overall quality. The resulting MW sub-sample comprises 25 galaxies, that are marked in Table \ref{tab:master} with an asterisk next to the name of the galaxy.

\begin{figure}
    \centering
    \begin{subfigure}{ \linewidth}
    \includegraphics[width=\linewidth]{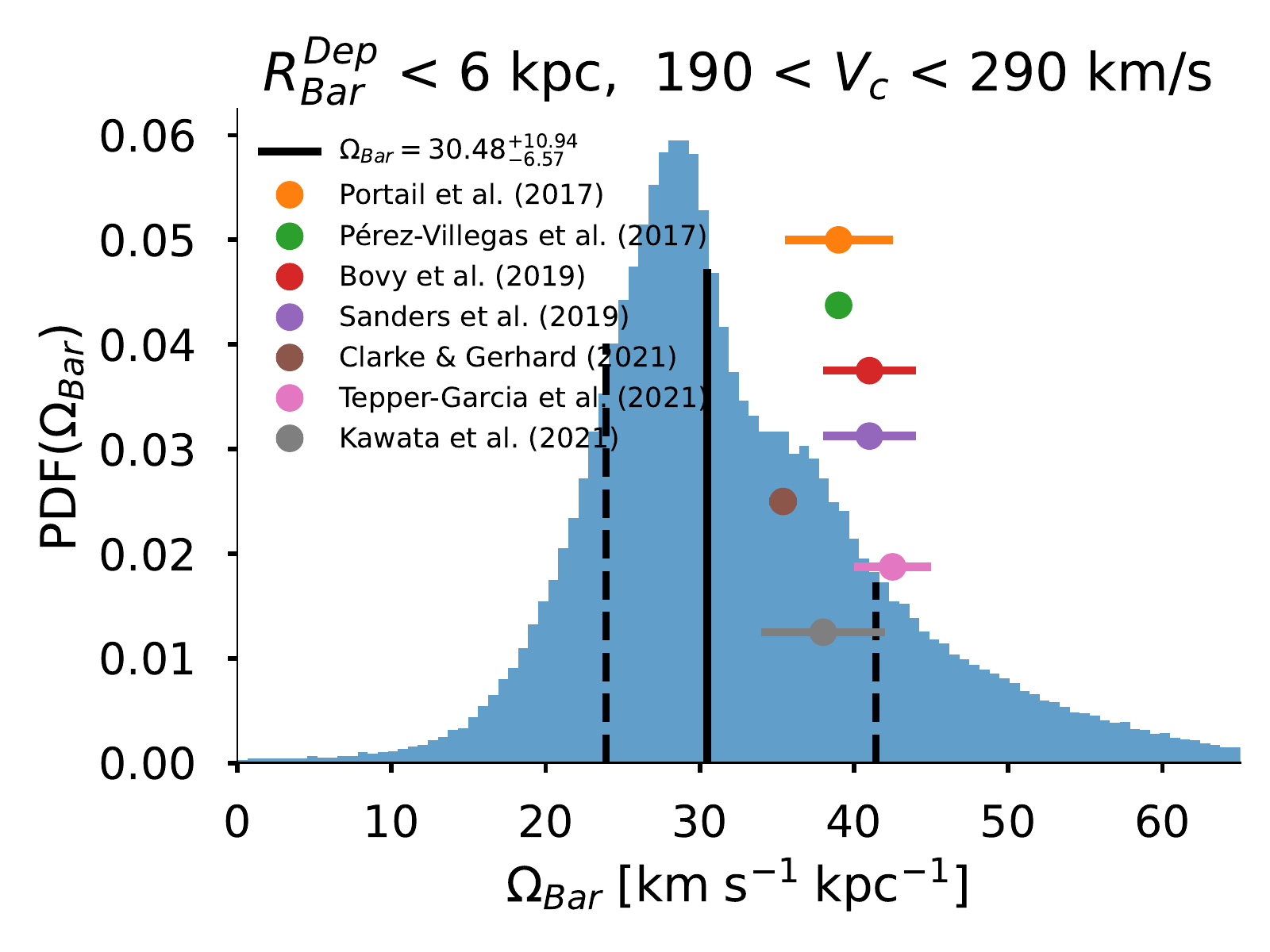}
    \end{subfigure}
    \begin{subfigure}{ \linewidth}
    \includegraphics[width=\linewidth]{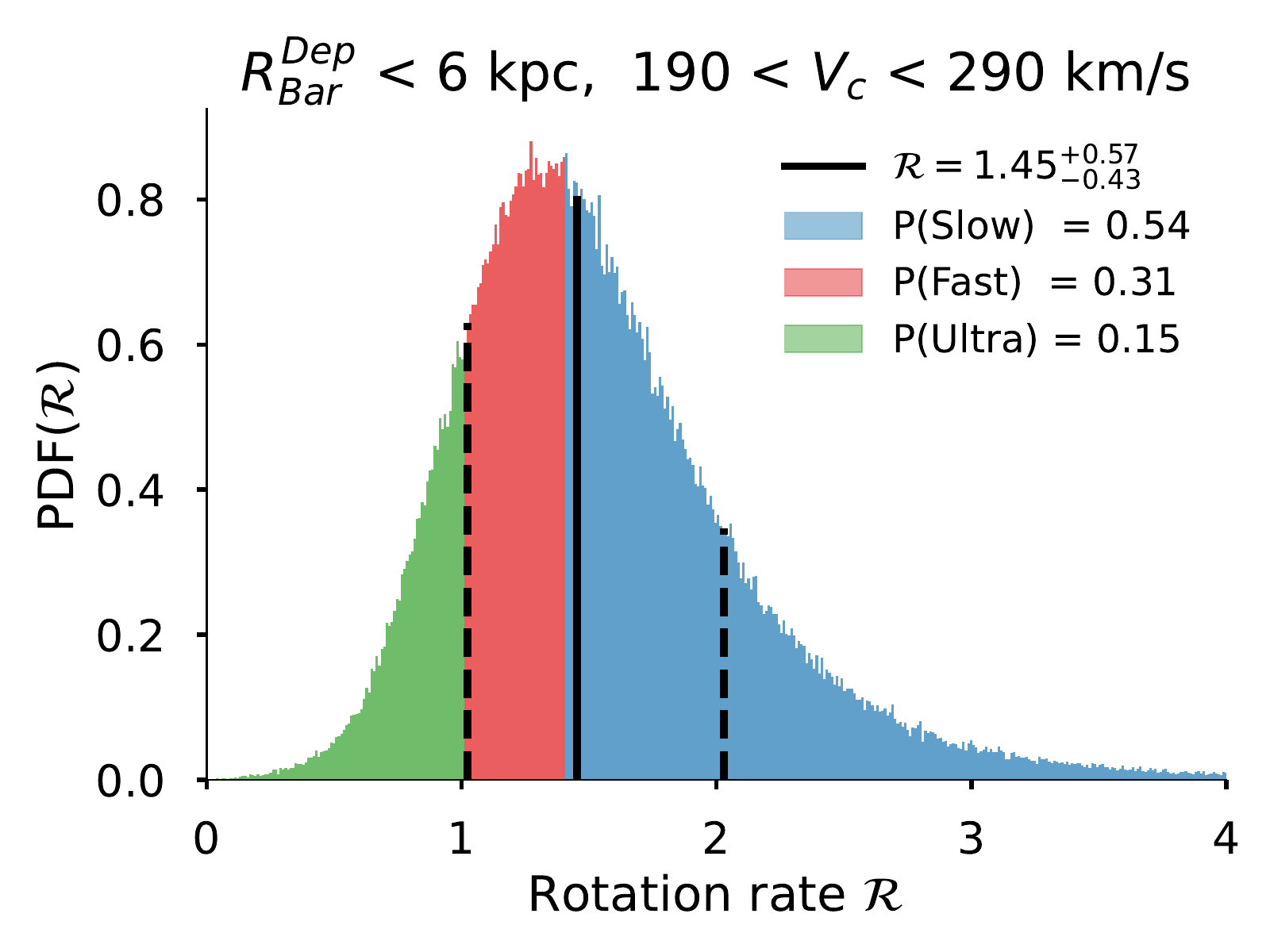}
    \end{subfigure}

    \caption{Bar pattern speed and \Rpar{} distributions of the MW sub-sample. This sub-sample is built by adding cuts in the bar lengths $(R_{bar}^{Dep}<6$kpc) and the disc circular velocity ($190 < V_c < 290$ km s$^{-1}$).}
    \label{fig:Omega_MW}
\end{figure}

Figure \ref{fig:Omega_MW} shows the resulting distribution in \Om{} and \Rpar{}. Most recent estimates of the bar pattern speed in our Galaxy are in the range $\Omega_{bar} \sim 35-45$ km s$^{-1}$ kpc$^{-1}$, which lies within the 1-sigma upper limit of our distribution. 

\section{Discussion}
\label{sec:discussion}


\subsection{Improving future measurements}
\label{sec:improving_measures}

Throughout this work, we have discussed many error sources and biases that appear when measuring the bar pattern speed and derived quantities. The most important of these is the disc PA, where a few degrees of error can change the measurement dramatically. This would not be a problem if different measurements agreed on their uncertainties, but this is usually not the case. In this work, we choose to do an equal weighting of different PA measurements (by increasing their errors artificially until they become consistent), and impose a condition over the linearity of the TW-integrals (see section \ref{sec:PA_measure}). An improvement could be made by using a criterion based on the goodness of fit $\chi^2$ (maybe with a threshold condition, or using $\chi^2$ to weight the PA), or by modelling the systemic errors of each measurement. 

Our procedure is biased towards measurements that produce a linear behaviour in the TW integrals (we are weighting our observations based on the method). Sometimes this includes galaxies that should not work with the TW-method, but nonetheless, produce a measurement that has physical sense (see our discussion in sections \ref{sec:where_are_the_ultrafast} and \ref{sec:mw_like}).

We estimated the geometric error of \Om{} using a MC procedure over the inclination, PA, centre and slits length. However, the number of slits could be an important parameter to consider. In most galaxies the TW integrals follow a linear trend up to near the bar end. We tried to use a number of slits to fill the area enclosed by $R_\epsilon$, however, there are exceptions. In some galaxies the linearity can only be seen in the innermost slits, while in others, the central slits are clustered together and only the outermost slits follow a lineal trend. By letting the number of slits be a free parameter (within the uncertainties of the bar radius) the measurements would capture some of this behaviour. 
 
Also important is the bar radius estimation, where morphological features (bulges, rings, ansae, spirals) affect the different measurement techniques. In this work, we choose to model the bar radius using a log-normal distribution with $R_\epsilon$ and $R_{PA}$ as the mean and 2-sigma upper limit of the distribution. This could be improved by incorporating other measurements like the the 2-D decomposition model of the surface brightness or the bar maximum torque.

Using hydrodynamical simulations, \cite{Hilmi2020} estimated that observations could be overestimating the bar radius between $\sim 15 - 55$ per cent, depending on the timescale and orientation of the galactic structures. \cite{Cuomo2021} proposed the maximum torque radius as a better estimator for the bar radius (hereafter $R_{Qb}$). However, this method comes other with important limitations and biases. Most importantly, in various galaxies $R_{Qb}$ does not agree with the visual estimate of what constitutes the bar. Also, the method is affected by the bulge-to-total ratio (B/T) as it can dilute $Q_b$ and increase $R_{Qb}$ \citep{Diaz2016_Bar_characteristics}. The B/T ratio also affects the isophote method, but only in exponential weak bars \citep{Lee2020}.

\begin{figure}
    \centering
    \includegraphics[width=\linewidth]{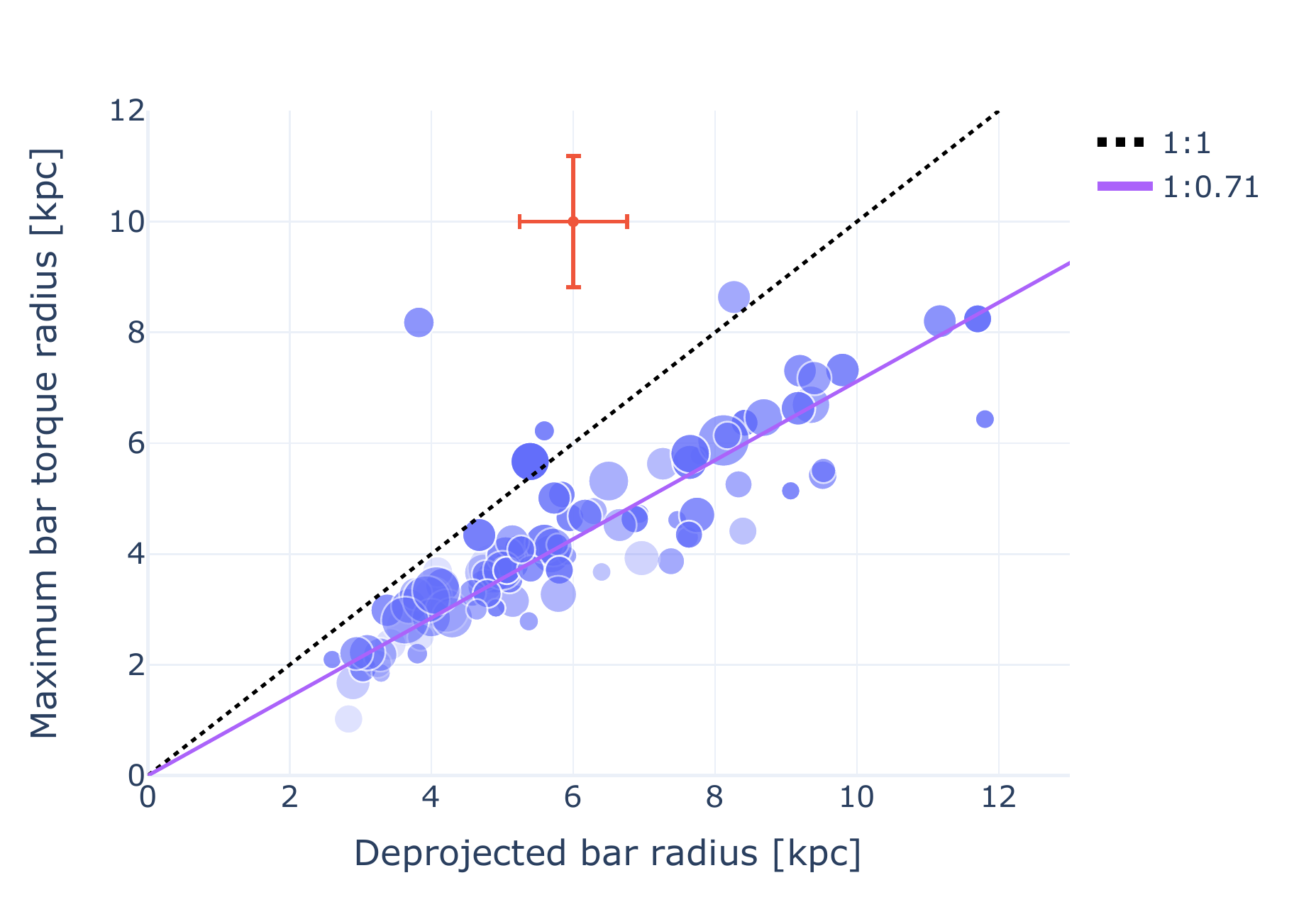}
    \caption{Bar radius estimated using the maximum torque versus the deprojected bar radius estimated from the maximum ellipticity isophote. }
    \label{fig:bar_med_rQ}
\end{figure}

For comparison purposes, we estimated $R_{Qb}$ by de-projecting the r-band DESI image and solving the Poisson equation assuming a constant mass-to-light ratio \citep{Buta2001}. In Figure \ref{fig:bar_med_rQ} we show both bar measuring methods in our sample. We found $R_{Qb}$ to be $\sim 30\%$ smaller than $R_\epsilon$ after deprojection.

We tried using $R_{Qb}$ as the 2-sigma lower limit of the distribution (with $R_\epsilon$ as the mean) but in some galaxies, the difference between $R_\epsilon$ and $R_{Qb}$ is so significant, that the distribution had a long tail towards high bar radii.  We got some reasonable results using both $R_{Qb}$ and $R_{PA}$ as the lower and upper 2-sigma limits, but the distribution of \Rpar{} of the complete sample widens significantly, including some ultra-fast (\Rpar{}$\sim 1$) and ultra-slow bars (\Rpar{}$\sim 4$).  Thus, we choose to use the combination that produced more consistent results. Maybe incorporating a Gaussian Mixture Model to estimate the intrinsic biases could help improve some of these measurements. 

\section{Conclusions}
\label{sec:conclusions}

\begin{itemize}
    \item  We have measured the bar pattern speed, bar radius, corotation radius and rotation rate of a sample of 97 MaNGA galaxies using the TW method. The sample was constructed to resemble the Milky-Way by using cuts in the stellar mass and the morphological type. The TW integrals were computed over the stellar component.

    \item We used 3 independent measurements of the disc PA from different sources: isophote fitting of the surface brightness, symmetrization of the stellar velocity field and fitting a kinematic model to the H$_\alpha$ velocity field. We assume all measurements are biased and increased their errors to produce an equal weighted PA. We penalise PAs when the TW integrals are not linear. 

    \item We used a MC procedure to estimate the \Om{} distribution by sampling the PA, inclination, centre and slit length. This procedure let us disentangle the effects of each error source.  

    \item We used the rotation curve from the kinematic bi-symmetric model of H$_\alpha$ velocity field· The model fits non-circular motions produced by the bar, and does not require an asymmetric drift correction. 

    \item Two measurements of the bar radius were obtained from the isophotal procedure $R_\epsilon$ and $R_{PA}$. We choose to model the bar radius with a log-normal distribution using these measurements as parameters (mean and 2-sigma).

    \item We found two significant correlations within our sample: (i)  \Om{}-$R_{bar}^{Dep}$-$\log M_* / M_{\odot}$ which relates the size of the system to the rotation frequency and (ii) \Om{}-$V_c$-\Rpar{}, that suggest fast rotating discs tend to host high pattern speed bars, but slow in rotation rate \Rpar{}.
    
    \item We also looked for correlations with various global galactic properties. We found a weak correlation between \Rpar{} and the gas fraction. Also, the weak relations between \Om{} and \Rpar{} with the Sérsic index and $v/\sigma$ ratio within 1 effective radius, suggest that more concentrated mass distributions produce bars that rotate at lower pattern speeds, but faster in \Rpar{}.

    \item We identify the inclination angle and the relative orientation of the disc-bar as possible sources of ultra-fast rotating bars in our sample. Using a cut in both parameters reduced the frequency of ultra-fast from 20\% to 10\% of the sample. 

    \item We build a sub-sample of MW galaxies using these quality cuts, and an additional cut in the bar radius and disc circular velocity. The most recent measurements of the MW bar pattern speed lie within the upper 1-sigma of our distribution. 

    \item We suggest future measurements to take into account all possible biases in the procedure or, if possible, model these biases. 

\end{itemize}

\section*{Data availability}

All Figures and relevant tables are available in the public repository \url{https://github.com/lgarma/MWA_pattern_speed}. Other data that support the findings of this study are available from the corresponding author, upon reasonable request.

\section*{Acknowledgements}

We thank the referee Isabel Perez for  the careful revision and useful comments that significantly improved the quality of the paper.
LGO acknowledge support from CONACyT scholarship. LGO and LMM acknowledge support from PAPIIT IA101520 and IA104022 grants.
OV akcnowledges support from PAPIIT grants: IG101620 and IG10122. 
EAO acknowledge support from the SECTEI (Secretaría de Educaci\'on, Ciencia, Tecnolog\'ia e Innovaci\'on de la Ciudad de M\'exico) under the Postdoctoral Fellowship SECTEI/170/2021 and CM-SECTEI/303/2021.




\bibliographystyle{mnras}
\bibliography{references}







\appendix
\section{Table}
\clearpage
\onecolumn
\begin{landscape} 
\renewcommand{\arraystretch}{1.4} 
\begin{longtable}{ccccccccccc} 
   \hline \hline 
    Galaxy & $\log(M/M_\odot)$ & Disc PA (weighted) & Bar PA & $i$ & $V_c$ & $R_{bar}$ & $R_{bar}^{Dep}$ & $\Omega_{bar}$ & $R_{cr}$ & $\mathcal{R}$  \\ 
     &  & [$\degree$] & [$\degree$] & [$\degree$] & [km s$^{-1}$] & [arcsec] & [kpc] & [km s$^{-1}$ kpc$^{-1}$] & [kpc] &   \\ 
    (1) & (2) & (3) & (4) & (5) & (6) & (7) & (8) & (9) & (10) & (11)  \\ 
   \hline 
7495-12704* & $10.8$  & $172.8 \pm 1.1$ (s+m)  & $147.8 \pm 3.5$ & $55.4^{+2.5}_{-2.6}$  & $223.1 \pm 1.2 $ & $5.1 \pm 2.0 $ & $3.9^{+1.3}_{-1.0}$  & $26.9^{+4.1}_{-5.0}$  & $7.6^{+2.1}_{-1.3}$  & $2.0^{+0.9}_{-0.6}$ \\ 
7958-6101 & $10.5$  & $153.9 \pm 0.7$ (p+s+m)  & $158.6 \pm 0.9$ & $40.7^{+1.1}_{-1.2}$  & $182.3 \pm 1.1 $ & $5.2 \pm 0.3 $ & $2.6^{+0.1}_{-0.1}$  & $47.9^{+13.1}_{-16.5}$  & $3.5^{+1.5}_{-0.9}$  & $1.4^{+0.6}_{-0.4}$ \\ 
7958-3702 & $10.7$  & $144.9 \pm 1.0$ (p+s)  & $68.2 \pm 4.2$ & $50.0^{+2.5}_{-2.6}$  & $246.8 \pm 1.7 $ & $2.3 \pm 0.2 $ & $2.8^{+0.3}_{-0.3}$  & $74.3^{+11.2}_{-11.8}$  & $3.3^{+0.6}_{-0.5}$  & $1.2^{+0.3}_{-0.2}$ \\ 
7977-9102 & $10.8$  & $113.0 \pm 1.0$ (m)  & $138.2 \pm 2.5$ & $49.1^{+3.2}_{-3.4}$  & $311.9 \pm 1.5 $ & $2.3 \pm 0.5 $ & $3.7^{+0.8}_{-0.6}$  & $41.1^{+5.2}_{-5.5}$  & $7.4^{+1.2}_{-0.9}$  & $2.0^{+0.5}_{-0.4}$ \\ 
7993-12704 & $11.2$  & $127.6 \pm 1.5$ (p+s)  & $26.2 \pm 2.2$ & $30.7^{+3.0}_{-3.3}$  & $252.7 \pm 5.5 $ & $11.1 \pm 1.8 $ & $11.2^{+1.7}_{-1.6}$  & $24.1^{+6.9}_{-6.2}$  & $10.6^{+3.5}_{-2.4}$  & $0.9^{+0.4}_{-0.2}$ \\ 
8078-12703 & $10.9$  & $4.5 \pm 0.9$ (p+s)  & $68.0 \pm 1.1$ & $38.5^{+3.2}_{-3.4}$  & $294.3 \pm 2.5 $ & $8.2 \pm 0.9 $ & $5.9^{+0.6}_{-0.6}$  & $42.6^{+8.5}_{-8.0}$  & $6.5^{+1.8}_{-1.4}$  & $1.1^{+0.3}_{-0.3}$ \\ 
8085-3704 & $10.7$  & $104.0 \pm 3.0$ (s)  & $108.1 \pm 3.0$ & $31.8^{+7.0}_{-8.8}$  & $196.4 \pm 12.6 $ & $5.1 \pm 1.2 $ & $4.0^{+0.8}_{-0.7}$  & $34.8^{+10.8}_{-8.0}$  & $4.5^{+1.8}_{-1.4}$  & $1.1^{+0.5}_{-0.4}$ \\ 
8088-3701* & $10.5$  & $44.0 \pm 2.0$ (s)  & $169.8 \pm 3.1$ & $30.3^{+5.7}_{-6.9}$  & $244.7 \pm 3.0 $ & $4.0 \pm 0.7 $ & $4.2^{+0.6}_{-0.6}$  & $31.4^{+13.4}_{-10.7}$  & $7.9^{+2.6}_{-1.9}$  & $1.9^{+0.7}_{-0.5}$ \\ 
8091-6101 & $11.2$  & $85.8 \pm 0.6$ (p+s+m)  & $42.8 \pm 2.7$ & $46.1^{+1.8}_{-1.9}$  & $259.3 \pm 3.2 $ & $7.3 \pm 0.8 $ & $7.6^{+0.8}_{-0.7}$  & $23.3^{+2.6}_{-3.0}$  & $11.1^{+1.6}_{-1.2}$  & $1.5^{+0.3}_{-0.2}$ \\ 
8091-12701 & $10.8$  & $162.1 \pm 0.8$ (s+m)  & $111.0 \pm 0.4$ & $37.6^{+4.2}_{-4.7}$  & $224.7 \pm 11.9 $ & $8.2 \pm 0.9 $ & $11.7^{+1.2}_{-1.2}$  & $11.8^{+1.7}_{-1.5}$  & $18.4^{+2.9}_{-2.4}$  & $1.6^{+0.3}_{-0.3}$ \\ 
8135-6103 & $11.0$  & $34.9 \pm 2.4$ (m)  & $8.3 \pm 2.2$ & $37.4^{+0.5}_{-0.5}$  & $239.8 \pm 4.4 $ & $7.5 \pm 1.2 $ & $8.3^{+1.2}_{-1.1}$  & $20.2^{+4.1}_{-5.1}$  & $11.0^{+3.0}_{-2.0}$  & $1.3^{+0.4}_{-0.3}$ \\ 
8146-9102 & $10.6$  & $17.0 \pm 1.0$ (s)  & $157.4 \pm 4.2$ & $54.4^{+2.7}_{-2.8}$  & $209.5 \pm 2.4 $ & $3.9 \pm 0.6 $ & $6.2^{+1.0}_{-0.9}$  & $23.1^{+1.8}_{-1.7}$  & $8.7^{+0.5}_{-0.4}$  & $1.4^{+0.2}_{-0.2}$ \\ 
8241-6102 & $10.6$  & $117.0 \pm 1.5$ (s)  & $174.3 \pm 5.3$ & $21.0^{+1.7}_{-1.8}$  & $297.6 \pm 2.3 $ & $4.0 \pm 0.2 $ & $3.3^{+0.2}_{-0.2}$  & $45.2^{+10.1}_{-8.5}$  & $5.9^{+1.3}_{-1.3}$  & $1.8^{+0.4}_{-0.4}$ \\ 
8245-12702 & $11.1$  & $17.7 \pm 0.7$ (p+s)  & $170.3 \pm 3.8$ & $51.8^{+2.7}_{-2.8}$  & $351.2 \pm 3.5 $ & $10.8 \pm 1.8 $ & $9.4^{+1.5}_{-1.3}$  & $26.3^{+1.9}_{-2.8}$  & $11.6^{+1.0}_{-0.7}$  & $1.3^{+0.2}_{-0.2}$ \\ 
8257-6103 & $11.2$  & $135.9 \pm 0.9$ (s+m)  & $54.6 \pm 4.4$ & $25.8^{+2.6}_{-2.9}$  & $426.1 \pm 5.9 $ & $4.2 \pm 0.4 $ & $6.3^{+0.6}_{-0.6}$  & $32.1^{+7.0}_{-5.9}$  & $11.9^{+2.2}_{-2.1}$  & $1.9^{+0.4}_{-0.4}$ \\ 
8312-12702* & $10.6$  & $92.5 \pm 1.4$ (p+s)  & $123.6 \pm 2.5$ & $38.9^{+4.2}_{-4.7}$  & $230.2 \pm 1.1 $ & $6.8 \pm 1.6 $ & $5.0^{+1.0}_{-0.9}$  & $29.2^{+4.1}_{-3.3}$  & $7.5^{+0.8}_{-0.8}$  & $1.5^{+0.3}_{-0.3}$ \\ 
8312-12705 & $10.5$  & $92.4 \pm 5.6$ (m)  & $130.8 \pm 6.5$ & $36.7^{+5.6}_{-6.4}$  & $108.2 \pm 11.9 $ & $7.4 \pm 1.7 $ & $5.4^{+1.1}_{-1.0}$  & $14.7^{+6.3}_{-5.1}$  & $5.6^{+4.8}_{-2.8}$  & $1.0^{+0.9}_{-0.5}$ \\ 
8319-12704 & $10.7$  & $109.2 \pm 1.0$ (p)  & $146.0 \pm 1.9$ & $42.8^{+1.6}_{-1.6}$  & $176.9 \pm 1.2 $ & $6.8 \pm 0.7 $ & $4.8^{+0.5}_{-0.4}$  & $30.9^{+2.5}_{-2.3}$  & $5.7^{+0.4}_{-0.4}$  & $1.2^{+0.1}_{-0.1}$ \\ 
8320-6101* & $10.4$  & $4.1 \pm 1.0$ (p+s)  & $60.2 \pm 7.0$ & $47.4^{+2.0}_{-2.0}$  & $203.2 \pm 12.2 $ & $4.2 \pm 0.7 $ & $3.3^{+0.5}_{-0.4}$  & $54.1^{+9.8}_{-10.5}$  & $3.1^{+1.0}_{-0.9}$  & $0.9^{+0.3}_{-0.3}$ \\ 
8324-12702* & $11.0$  & $68.5 \pm 0.8$ (p+m)  & $23.9 \pm 7.5$ & $35.9^{+1.2}_{-1.3}$  & $252.4 \pm 1.1 $ & $7.5 \pm 0.4 $ & $5.4^{+0.3}_{-0.3}$  & $33.3^{+7.0}_{-7.4}$  & $7.2^{+2.1}_{-1.5}$  & $1.3^{+0.4}_{-0.3}$ \\ 
8341-12704 & $10.7$  & $55.5 \pm 1.9$ (p+m)  & $80.9 \pm 1.2$ & $25.2^{+3.3}_{-3.7}$  & $116.7 \pm 1.4 $ & $7.5 \pm 0.4 $ & $4.9^{+0.2}_{-0.2}$  & $18.5^{+4.6}_{-4.1}$  & $5.4^{+1.8}_{-1.6}$  & $1.1^{+0.4}_{-0.3}$ \\ 
8444-12703* & $10.9$  & $4.9 \pm 0.6$ (s+m)  & $31.7 \pm 9.5$ & $37.2^{+1.2}_{-1.2}$  & $237.7 \pm 1.3 $ & $5.7 \pm 0.9 $ & $5.0^{+0.7}_{-0.7}$  & $40.9^{+4.0}_{-3.9}$  & $4.5^{+0.8}_{-0.8}$  & $0.9^{+0.2}_{-0.2}$ \\ 
8450-9102* & $10.4$  & $14.7 \pm 1.1$ (p+s+m)  & $-6.6 \pm 4.4$ & $48.6^{+1.4}_{-1.4}$  & $199.0 \pm 2.2 $ & $3.9 \pm 0.4 $ & $4.6^{+0.4}_{-0.4}$  & $29.5^{+1.7}_{-1.9}$  & $6.6^{+0.4}_{-0.4}$  & $1.4^{+0.2}_{-0.2}$ \\ 
8453-12701 & $10.4$  & $109.0 \pm 1.3$ (s)  & $45.6 \pm 2.8$ & $41.6^{+2.3}_{-2.4}$  & $149.5 \pm 4.0 $ & $5.6 \pm 0.3 $ & $3.8^{+0.2}_{-0.2}$  & $23.4^{+4.5}_{-5.4}$  & $4.8^{+2.0}_{-1.5}$  & $1.3^{+0.5}_{-0.4}$ \\ 
8454-12702 & $11.0$  & $66.0 \pm 1.7$ (s)  & $93.1 \pm 2.8$ & $45.4^{+2.7}_{-2.8}$  & $250.9 \pm 2.5 $ & $4.4 \pm 0.5 $ & $8.3^{+0.8}_{-0.8}$  & $26.1^{+2.1}_{-1.8}$  & $9.9^{+0.7}_{-1.0}$  & $1.2^{+0.2}_{-0.2}$ \\ 
8465-12705 & $10.9$  & $49.3 \pm 0.7$ (s+m)  & $34.9 \pm 4.5$ & $53.5^{+4.8}_{-5.1}$  & $290.6 \pm 2.1 $ & $5.6 \pm 0.6 $ & $3.8^{+0.5}_{-0.4}$  & $65.7^{+10.2}_{-10.2}$  & $4.9^{+0.3}_{-0.3}$  & $1.3^{+0.2}_{-0.2}$ \\ 
8486-6101* & $10.8$  & $114.9 \pm 1.1$ (s+m)  & $77.6 \pm 3.0$ & $40.4^{+2.8}_{-3.0}$  & $192.7 \pm 3.9 $ & $2.9 \pm 0.9 $ & $4.2^{+1.1}_{-0.9}$  & $27.5^{+3.0}_{-2.9}$  & $6.8^{+0.8}_{-0.7}$  & $1.6^{+0.5}_{-0.4}$ \\ 
8552-9101* & $11.0$  & $110.2 \pm 0.9$ (s+m)  & $135.4 \pm 1.7$ & $32.7^{+1.3}_{-1.3}$  & $203.8 \pm 1.7 $ & $3.9 \pm 0.4 $ & $5.8^{+0.5}_{-0.5}$  & $26.7^{+2.7}_{-3.0}$  & $7.1^{+0.7}_{-0.6}$  & $1.2^{+0.2}_{-0.2}$ \\ 
8561-3704 & $10.4$  & $169.2 \pm 1.4$ (p+s)  & $156.5 \pm 3.3$ & $39.5^{+2.0}_{-2.0}$  & $234.3 \pm 23.3 $ & $5.6 \pm 1.3 $ & $5.7^{+1.1}_{-1.0}$  & $27.0^{+3.6}_{-4.8}$  & $4.9^{+2.6}_{-2.2}$  & $0.8^{+0.5}_{-0.4}$ \\ 
8589-12705 & $11.1$  & $167.8 \pm 1.3$ (p+m)  & $34.7 \pm 2.7$ & $25.0^{+2.3}_{-2.5}$  & $309.2 \pm 1.3 $ & $7.4 \pm 1.2 $ & $4.8^{+0.7}_{-0.7}$  & $28.9^{+9.1}_{-8.8}$  & $10.1^{+3.4}_{-2.3}$  & $2.1^{+0.8}_{-0.5}$ \\ 
8596-12704 & $10.9$  & $82.6 \pm 1.8$ (p+m)  & $58.4 \pm 1.9$ & $45.0^{+3.0}_{-3.2}$  & $258.2 \pm 1.5 $ & $7.6 \pm 1.3 $ & $6.7^{+1.0}_{-0.9}$  & $25.6^{+4.4}_{-4.7}$  & $9.5^{+1.8}_{-1.4}$  & $1.4^{+0.3}_{-0.3}$ \\ 
8597-12703* & $10.8$  & $170.4 \pm 1.9$ (p+s+m)  & $137.3 \pm 1.6$ & $42.6^{+1.1}_{-1.1}$  & $228.2 \pm 0.9 $ & $8.2 \pm 0.9 $ & $5.1^{+0.5}_{-0.5}$  & $28.4^{+5.5}_{-6.8}$  & $7.4^{+1.9}_{-1.2}$  & $1.5^{+0.4}_{-0.3}$ \\ 
8602-3701 & $10.3$  & $153.1 \pm 0.7$ (p+s)  & $186.3 \pm 3.6$ & $50.0^{+1.6}_{-1.7}$  & $175.4 \pm 3.1 $ & $4.0 \pm 0.4 $ & $3.2^{+0.3}_{-0.3}$  & $26.9^{+2.9}_{-3.5}$  & $5.8^{+0.7}_{-0.6}$  & $1.8^{+0.3}_{-0.3}$ \\ 
8602-12701 & $11.3$  & $156.2 \pm 2.2$ (p)  & $193.7 \pm 0.7$ & $39.8^{+1.8}_{-1.9}$  & $256.7 \pm 7.3 $ & $12.3 \pm 0.6 $ & $7.8^{+0.4}_{-0.4}$  & $24.7^{+6.8}_{-8.0}$  & $9.4^{+4.3}_{-2.6}$  & $1.2^{+0.5}_{-0.3}$ \\ 
8602-12705 & $11.2$  & $143.5 \pm 1.5$ (m)  & $185.5 \pm 0.7$ & $37.9^{+2.5}_{-2.6}$  & $184.4 \pm 4.7 $ & $11.1 \pm 1.2 $ & $8.4^{+0.8}_{-0.8}$  & $15.9^{+3.1}_{-2.8}$  & $11.4^{+2.5}_{-1.9}$  & $1.4^{+0.3}_{-0.3}$ \\ 
8612-12702 & $11.2$  & $47.8 \pm 2.1$ (p+s)  & $69.5 \pm 2.5$ & $47.5^{+3.9}_{-4.2}$  & $287.8 \pm 0.7 $ & $5.1 \pm 0.3 $ & $7.6^{+0.5}_{-0.4}$  & $20.6^{+5.8}_{-6.7}$  & $12.6^{+5.5}_{-3.9}$  & $1.6^{+0.7}_{-0.5}$ \\ 
8615-3701 & $10.8$  & $106.7 \pm 1.7$ (s+m)  & $72.2 \pm 3.4$ & $33.8^{+4.6}_{-5.2}$  & $137.9 \pm 3.5 $ & $2.5 \pm 1.0 $ & $3.6^{+1.2}_{-0.9}$  & $29.7^{+5.5}_{-4.3}$  & $3.7^{+0.9}_{-0.9}$  & $1.0^{+0.4}_{-0.3}$ \\ 
8616-6104 & $10.7$  & $32.0 \pm 0.7$ (p+s)  & $-0.6 \pm 6.4$ & $61.6^{+4.5}_{-4.7}$  & $251.7 \pm 6.7 $ & $3.9 \pm 0.4 $ & $7.0^{+1.3}_{-1.0}$  & $24.1^{+3.4}_{-4.1}$  & $9.8^{+1.9}_{-1.5}$  & $1.4^{+0.4}_{-0.3}$ \\ 
8622-12704 & $11.1$  & $76.6 \pm 1.3$ (p+m)  & $62.8 \pm 0.2$ & $42.8^{+1.9}_{-1.9}$  & $265.3 \pm 4.8 $ & $3.8 \pm 0.2 $ & $5.9^{+0.3}_{-0.3}$  & $27.3^{+6.3}_{-8.9}$  & $8.5^{+3.6}_{-2.0}$  & $1.4^{+0.6}_{-0.3}$ \\ 
8624-9102 & $10.6$  & $142.0 \pm 1.3$ (s)  & $132.4 \pm 1.9$ & $46.6^{+2.0}_{-2.1}$  & $198.4 \pm 3.5 $ & $8.9 \pm 0.9 $ & $5.4^{+0.5}_{-0.5}$  & $25.3^{+5.1}_{-6.8}$  & $7.1^{+1.8}_{-1.1}$  & $1.3^{+0.4}_{-0.2}$ \\ 
8625-12703 & $11.0$  & $88.0 \pm 1.0$ (s)  & $129.2 \pm 5.8$ & $69.1^{+0.4}_{-0.4}$  & $266.8 \pm 3.7 $ & $5.0 \pm 0.5 $ & $7.4^{+0.7}_{-0.7}$  & $27.6^{+6.3}_{-8.3}$  & $7.7^{+3.5}_{-1.9}$  & $1.0^{+0.5}_{-0.3}$ \\ 
8655-3701 & $11.0$  & $148.0 \pm 1.7$ (s)  & $160.0 \pm 2.0$ & $36.0^{+6.7}_{-8.1}$  & $140.8 \pm 3.2 $ & $5.9 \pm 1.0 $ & $9.2^{+1.4}_{-1.2}$  & $15.4^{+5.2}_{-4.9}$  & $8.7^{+3.4}_{-2.4}$  & $1.0^{+0.4}_{-0.3}$ \\ 
8656-6103* & $11.2$  & $14.6 \pm 1.6$ (s+m)  & $-39.6 \pm 2.6$ & $27.4^{+5.0}_{-6.1}$  & $254.3 \pm 3.7 $ & $3.3 \pm 0.6 $ & $4.8^{+0.7}_{-0.6}$  & $32.7^{+11.0}_{-8.6}$  & $7.9^{+2.5}_{-1.8}$  & $1.7^{+0.6}_{-0.4}$ \\ 
8713-9102 & $10.6$  & $151.5 \pm 1.1$ (s+m)  & $187.7 \pm 3.1$ & $34.6^{+4.2}_{-4.7}$  & $131.9 \pm 26.2 $ & $6.8 \pm 0.7 $ & $5.3^{+0.5}_{-0.5}$  & $21.1^{+4.0}_{-3.3}$  & $5.3^{+1.6}_{-1.3}$  & $1.0^{+0.3}_{-0.3}$ \\ 
8715-12701* & $10.9$  & $134.0 \pm 1.3$ (s)  & $166.5 \pm 1.6$ & $40.2^{+1.1}_{-1.1}$  & $257.7 \pm 4.2 $ & $10.3 \pm 1.7 $ & $5.7^{+0.8}_{-0.8}$  & $25.7^{+3.8}_{-5.4}$  & $7.8^{+3.4}_{-1.9}$  & $1.4^{+0.6}_{-0.4}$ \\ 
8718-12701 & $10.7$  & $72.1 \pm 2.4$ (m)  & $88.2 \pm 0.8$ & $37.0^{+1.9}_{-1.9}$  & $132.3 \pm 3.3 $ & $7.5 \pm 0.8 $ & $8.2^{+0.8}_{-0.7}$  & $10.0^{+1.6}_{-2.2}$  & $12.3^{+3.6}_{-2.0}$  & $1.5^{+0.5}_{-0.3}$ \\ 
8721-6103 & $11.2$  & $50.8 \pm 0.8$ (s+m)  & $136.3 \pm 3.0$ & $32.7^{+5.4}_{-6.4}$  & $352.8 \pm 3.5 $ & $4.5 \pm 0.7 $ & $5.1^{+0.8}_{-0.7}$  & $40.6^{+13.1}_{-9.3}$  & $8.5^{+2.3}_{-2.2}$  & $1.6^{+0.5}_{-0.5}$ \\ 
8938-12702 & $11.1$  & $53.4 \pm 1.4$ (s+m)  & $32.1 \pm 0.3$ & $43.9^{+1.4}_{-1.4}$  & $303.4 \pm 2.5 $ & $7.6 \pm 0.4 $ & $6.9^{+0.3}_{-0.3}$  & $25.9^{+8.2}_{-9.2}$  & $10.8^{+5.2}_{-3.3}$  & $1.5^{+0.7}_{-0.5}$ \\ 
8940-12702* & $11.0$  & $140.2 \pm 0.9$ (p+m)  & $112.2 \pm 1.3$ & $44.7^{+1.0}_{-1.1}$  & $240.1 \pm 3.2 $ & $8.9 \pm 2.1 $ & $5.7^{+1.1}_{-1.0}$  & $36.7^{+2.2}_{-2.4}$  & $5.7^{+0.4}_{-0.4}$  & $1.0^{+0.2}_{-0.2}$ \\ 
8948-12702 & $11.0$  & $1.0 \pm 1.3$ (s)  & $49.7 \pm 0.6$ & $19.8^{+1.1}_{-1.2}$  & $246.5 \pm 5.7 $ & $6.0 \pm 1.0 $ & $3.4^{+0.5}_{-0.5}$  & $46.5^{+28.5}_{-21.1}$  & $4.1^{+1.2}_{-0.5}$  & $1.3^{+0.4}_{-0.3}$ \\ 
8978-9101* & $10.7$  & $93.6 \pm 0.7$ (s+m)  & $251.5 \pm 0.8$ & $35.1^{+2.7}_{-2.9}$  & $202.1 \pm 2.3 $ & $5.8 \pm 0.6 $ & $3.8^{+0.4}_{-0.3}$  & $26.6^{+3.6}_{-4.0}$  & $6.4^{+1.2}_{-0.9}$  & $1.7^{+0.4}_{-0.3}$ \\ 
8978-3701 & $10.4$  & $19.6 \pm 1.4$ (p+s)  & $126.0 \pm 0.5$ & $46.4^{+5.1}_{-5.6}$  & $245.1 \pm 47.7 $ & $5.6 \pm 0.3 $ & $4.8^{+0.5}_{-0.5}$  & $32.6^{+13.0}_{-11.2}$  & $5.8^{+2.3}_{-1.9}$  & $1.2^{+0.5}_{-0.4}$ \\ 
8979-12701 & $11.2$  & $121.0 \pm 1.1$ (s+m)  & $161.9 \pm 4.4$ & $50.6^{+0.7}_{-0.7}$  & $286.5 \pm 1.2 $ & $4.2 \pm 1.0 $ & $9.3^{+1.9}_{-1.6}$  & $24.7^{+2.8}_{-2.5}$  & $10.4^{+1.3}_{-1.3}$  & $1.1^{+0.3}_{-0.2}$ \\ 
8983-12701* & $10.5$  & $70.4 \pm 0.8$ (p+s+m)  & $125.1 \pm 8.7$ & $51.6^{+4.1}_{-4.4}$  & $245.6 \pm 7.4 $ & $3.8 \pm 0.6 $ & $3.1^{+0.6}_{-0.5}$  & $46.6^{+9.0}_{-9.2}$  & $4.8^{+1.0}_{-1.0}$  & $1.5^{+0.4}_{-0.4}$ \\ 
8983-3703* & $10.8$  & $174.9 \pm 0.6$ (p+s+m)  & $147.3 \pm 3.5$ & $57.9^{+3.3}_{-3.5}$  & $205.1 \pm 2.7 $ & $4.9 \pm 1.9 $ & $4.1^{+1.3}_{-1.1}$  & $29.2^{+4.9}_{-6.6}$  & $6.9^{+2.0}_{-1.5}$  & $1.7^{+0.8}_{-0.5}$ \\ 
8984-12704* & $10.4$  & $114.0 \pm 1.7$ (s)  & $73.7 \pm 0.8$ & $35.1^{+6.0}_{-7.1}$  & $255.3 \pm 5.0 $ & $9.3 \pm 1.0 $ & $5.8^{+0.6}_{-0.6}$  & $21.3^{+5.2}_{-4.0}$  & $11.4^{+1.9}_{-1.6}$  & $2.0^{+0.4}_{-0.3}$ \\ 
8985-9102 & $11.0$  & $127.0 \pm 1.7$ (m)  & $45.4 \pm 3.4$ & $57.5^{+1.6}_{-1.6}$  & $295.0 \pm 7.6 $ & $3.2 \pm 0.3 $ & $8.4^{+0.9}_{-0.8}$  & $23.1^{+5.5}_{-9.1}$  & $10.3^{+7.2}_{-4.2}$  & $1.2^{+0.9}_{-0.5}$ \\ 
8989-3703 & $10.6$  & $36.1 \pm 0.9$ (s+m)  & $47.3 \pm 0.5$ & $46.3^{+1.2}_{-1.3}$  & $363.2 \pm 20.4 $ & $8.2 \pm 1.4 $ & $4.7^{+0.7}_{-0.6}$  & $46.3^{+4.7}_{-5.4}$  & $4.8^{+1.5}_{-1.2}$  & $1.0^{+0.4}_{-0.3}$ \\ 
8993-12701 & $11.1$  & $138.0 \pm 0.9$ (s+m)  & $70.6 \pm 1.4$ & $54.9^{+7.4}_{-8.2}$  & $307.3 \pm 1.1 $ & $8.3 \pm 0.9 $ & $6.5^{+1.6}_{-1.1}$  & $53.0^{+14.3}_{-15.5}$  & $5.7^{+1.5}_{-0.9}$  & $0.9^{+0.3}_{-0.2}$ \\ 
9025-3703 & $11.1$  & $52.0 \pm 0.7$ (s+m)  & $163.1 \pm 2.2$ & $32.6^{+2.6}_{-2.8}$  & $303.5 \pm 17.6 $ & $4.3 \pm 0.2 $ & $5.6^{+0.3}_{-0.3}$  & $32.2^{+3.3}_{-3.0}$  & $7.5^{+0.6}_{-0.8}$  & $1.3^{+0.1}_{-0.1}$ \\ 
9028-12704* & $10.7$  & $55.6 \pm 0.9$ (p+s+m)  & $1.2 \pm 0.3$ & $49.3^{+1.9}_{-2.0}$  & $239.9 \pm 0.9 $ & $4.7 \pm 0.2 $ & $4.6^{+0.3}_{-0.3}$  & $35.7^{+10.2}_{-9.7}$  & $6.5^{+2.3}_{-1.6}$  & $1.4^{+0.5}_{-0.3}$ \\ 
9029-12704* & $10.8$  & $116.0 \pm 1.0$ (s)  & $85.0 \pm 2.0$ & $51.2^{+2.5}_{-2.6}$  & $227.6 \pm 0.8 $ & $6.3 \pm 1.0 $ & $5.1^{+0.8}_{-0.7}$  & $27.7^{+3.2}_{-3.6}$  & $8.0^{+1.1}_{-1.1}$  & $1.5^{+0.3}_{-0.3}$ \\ 
9042-12703 & $10.8$  & $147.0 \pm 1.1$ (s+m)  & $105.5 \pm 2.4$ & $40.1^{+8.8}_{-10.8}$  & $334.6 \pm 12.8 $ & $8.6 \pm 0.4 $ & $6.9^{+0.8}_{-0.6}$  & $36.9^{+10.0}_{-5.8}$  & $9.1^{+1.9}_{-1.7}$  & $1.3^{+0.3}_{-0.3}$ \\ 
9046-12702* & $10.9$  & $41.1 \pm 1.2$ (p+s+m)  & $89.5 \pm 1.7$ & $43.2^{+0.8}_{-0.8}$  & $245.2 \pm 1.5 $ & $6.1 \pm 0.6 $ & $5.1^{+0.5}_{-0.5}$  & $31.0^{+3.8}_{-4.0}$  & $7.4^{+0.8}_{-1.1}$  & $1.4^{+0.2}_{-0.2}$ \\ 
9047-12703 & $11.3$  & $104.7 \pm 0.7$ (p+s)  & $133.9 \pm 1.6$ & $54.1^{+0.6}_{-0.6}$  & $310.9 \pm 2.5 $ & $9.3 \pm 0.5 $ & $11.8^{+0.6}_{-0.6}$  & $21.5^{+1.6}_{-1.8}$  & $13.5^{+1.2}_{-0.9}$  & $1.1^{+0.1}_{-0.1}$ \\ 
9182-12703 & $11.1$  & $121.5 \pm 0.9$ (s+m)  & $133.4 \pm 1.4$ & $31.5^{+5.0}_{-5.8}$  & $278.0 \pm 3.1 $ & $5.8 \pm 1.3 $ & $8.7^{+1.8}_{-1.5}$  & $32.0^{+6.5}_{-4.5}$  & $7.8^{+1.8}_{-1.2}$  & $0.9^{+0.3}_{-0.2}$ \\ 
9187-3701 & $11.2$  & $143.9 \pm 2.1$ (s+m)  & $99.9 \pm 0.5$ & $42.0^{+5.5}_{-6.2}$  & $105.2 \pm 1.5 $ & $5.3 \pm 0.3 $ & $9.5^{+0.8}_{-0.7}$  & $8.2^{+2.3}_{-1.9}$  & $11.4^{+2.5}_{-2.1}$  & $1.2^{+0.3}_{-0.2}$ \\ 
9187-12704 & $10.9$  & $78.2 \pm 0.9$ (s+m)  & $93.7 \pm 1.7$ & $42.0^{+1.8}_{-1.8}$  & $214.5 \pm 1.3 $ & $8.4 \pm 0.9 $ & $5.2^{+0.5}_{-0.5}$  & $12.4^{+3.5}_{-3.7}$  & $14.3^{+3.7}_{-2.7}$  & $2.7^{+0.8}_{-0.5}$ \\ 
9196-12701 & $11.0$  & $123.9 \pm 2.2$ (s+m)  & $178.7 \pm 0.7$ & $29.9^{+0.5}_{-0.5}$  & $305.9 \pm 1.1 $ & $12.6 \pm 0.6 $ & $9.1^{+0.4}_{-0.4}$  & $23.8^{+6.9}_{-7.0}$  & $11.7^{+3.2}_{-2.2}$  & $1.3^{+0.4}_{-0.3}$ \\ 
9484-12703 & $10.7$  & $118.0 \pm 1.7$ (s)  & $101.5 \pm 5.9$ & $27.4^{+2.0}_{-2.1}$  & $277.4 \pm 5.8 $ & $9.5 \pm 0.5 $ & $6.4^{+0.3}_{-0.3}$  & $32.6^{+5.5}_{-5.7}$  & $5.4^{+2.5}_{-2.1}$  & $0.8^{+0.4}_{-0.3}$ \\ 
9490-6102 & $10.4$  & $165.4 \pm 1.0$ (p)  & $149.5 \pm 3.0$ & $58.3^{+4.1}_{-4.3}$  & $212.1 \pm 4.2 $ & $8.0 \pm 1.3 $ & $5.8^{+1.1}_{-0.9}$  & $27.9^{+4.9}_{-6.5}$  & $6.7^{+2.2}_{-1.3}$  & $1.2^{+0.5}_{-0.3}$ \\ 
9492-6101 & $10.4$  & $6.0 \pm 1.5$ (s)  & $158.4 \pm 1.9$ & $35.5^{+5.1}_{-5.8}$  & $138.6 \pm 4.4 $ & $4.6 \pm 0.8 $ & $2.9^{+0.4}_{-0.4}$  & $24.9^{+5.2}_{-4.4}$  & $3.7^{+1.5}_{-1.4}$  & $1.2^{+0.6}_{-0.5}$ \\ 
9502-12703* & $11.2$  & $22.0 \pm 1.3$ (s)  & $166.1 \pm 3.3$ & $42.0^{+7.9}_{-9.4}$  & $245.2 \pm 2.3 $ & $4.9 \pm 0.5 $ & $3.4^{+0.5}_{-0.4}$  & $29.3^{+7.7}_{-5.6}$  & $7.9^{+1.5}_{-1.4}$  & $2.3^{+0.5}_{-0.5}$ \\ 
9867-12704* & $10.5$  & $90.1 \pm 1.6$ (p+s)  & $56.2 \pm 2.7$ & $51.0^{+3.0}_{-3.1}$  & $202.5 \pm 1.8 $ & $3.5 \pm 0.6 $ & $2.9^{+0.5}_{-0.4}$  & $43.5^{+6.4}_{-5.9}$  & $3.4^{+1.6}_{-1.6}$  & $1.2^{+0.6}_{-0.5}$ \\ 
9881-12705 & $10.8$  & $49.2 \pm 1.3$ (p+s+m)  & $62.3 \pm 0.9$ & $36.0^{+3.3}_{-3.6}$  & $225.4 \pm 3.6 $ & $11.4 \pm 0.6 $ & $7.5^{+0.3}_{-0.4}$  & $22.2^{+2.9}_{-3.0}$  & $9.6^{+1.5}_{-1.2}$  & $1.3^{+0.2}_{-0.2}$ \\ 
9890-12702 & $11.2$  & $139.3 \pm 0.7$ (s+m)  & $121.6 \pm 1.8$ & $42.2^{+0.7}_{-0.7}$  & $240.6 \pm 0.9 $ & $5.3 \pm 1.2 $ & $7.7^{+1.5}_{-1.3}$  & $27.0^{+1.8}_{-2.0}$  & $8.5^{+0.8}_{-0.7}$  & $1.1^{+0.2}_{-0.2}$ \\ 
9894-12702 & $10.7$  & $6.9 \pm 1.1$ (p+s)  & $142.3 \pm 3.1$ & $43.9^{+10.2}_{-12.6}$  & $421.7 \pm 35.4 $ & $3.2 \pm 0.3 $ & $4.7^{+1.0}_{-0.7}$  & $39.6^{+10.7}_{-6.2}$  & $10.2^{+2.2}_{-2.0}$  & $2.1^{+0.6}_{-0.5}$ \\ 
10001-6102 & $10.7$  & $55.2 \pm 1.1$ (p)  & $72.9 \pm 2.5$ & $55.3^{+7.0}_{-7.7}$  & $296.9 \pm 27.4 $ & $6.8 \pm 1.6 $ & $4.3^{+1.0}_{-0.8}$  & $38.7^{+6.4}_{-7.0}$  & $8.4^{+3.9}_{-3.5}$  & $1.9^{+1.0}_{-0.8}$ \\ 
10213-12705* & $10.7$  & $152.2 \pm 2.2$ (p+s)  & $14.1 \pm 1.4$ & $40.6^{+4.2}_{-4.6}$  & $281.7 \pm 4.3 $ & $7.5 \pm 0.4 $ & $5.8^{+0.4}_{-0.4}$  & $24.1^{+4.4}_{-4.4}$  & $10.6^{+1.6}_{-1.3}$  & $1.8^{+0.3}_{-0.2}$ \\ 
10222-12704 & $11.1$  & $148.7 \pm 0.7$ (s+m)  & $38.7 \pm 14.7$ & $23.8^{+2.3}_{-2.6}$  & $346.4 \pm 2.7 $ & $4.4 \pm 0.7 $ & $7.2^{+1.1}_{-0.9}$  & $31.3^{+4.5}_{-3.9}$  & $10.5^{+1.6}_{-1.4}$  & $1.4^{+0.3}_{-0.3}$ \\ 
10510-6101 & $10.7$  & $36.1 \pm 2.4$ (p+s)  & $73.1 \pm 3.4$ & $43.2^{+7.8}_{-9.2}$  & $314.4 \pm 6.4 $ & $3.8 \pm 1.5 $ & $4.9^{+1.7}_{-1.3}$  & $42.3^{+11.9}_{-9.1}$  & $7.0^{+2.0}_{-1.7}$  & $1.4^{+0.7}_{-0.5}$ \\ 
10518-9102 & $10.5$  & $87.0 \pm 1.0$ (s)  & $73.5 \pm 0.5$ & $41.8^{+2.6}_{-2.7}$  & $197.9 \pm 9.1 $ & $8.2 \pm 0.4 $ & $5.0^{+0.2}_{-0.2}$  & $28.1^{+2.5}_{-2.6}$  & $6.1^{+0.7}_{-0.7}$  & $1.2^{+0.1}_{-0.1}$ \\ 
10520-6101 & $10.4$  & $177.3 \pm 2.9$ (p+s+m)  & $59.0 \pm 3.4$ & $23.5^{+2.9}_{-3.2}$  & $217.4 \pm 14.5 $ & $5.6 \pm 0.9 $ & $3.8^{+0.6}_{-0.5}$  & $42.6^{+10.2}_{-7.4}$  & $4.1^{+1.5}_{-1.6}$  & $1.1^{+0.5}_{-0.4}$ \\ 
11016-12703 & $11.2$  & $92.7 \pm 0.9$ (p+s+m)  & $64.4 \pm 5.3$ & $47.3^{+4.6}_{-5.0}$  & $309.1 \pm 28.9 $ & $6.6 \pm 1.1 $ & $7.6^{+1.2}_{-1.1}$  & $31.7^{+3.6}_{-3.4}$  & $9.4^{+1.4}_{-1.3}$  & $1.2^{+0.3}_{-0.2}$ \\ 
11017-12704* & $10.6$  & $66.6 \pm 1.4$ (p+s)  & $88.7 \pm 6.1$ & $47.2^{+7.2}_{-8.2}$  & $198.9 \pm 1.0 $ & $6.2 \pm 1.0 $ & $3.7^{+0.6}_{-0.5}$  & $44.7^{+11.2}_{-10.1}$  & $3.8^{+1.4}_{-1.1}$  & $1.0^{+0.4}_{-0.3}$ \\ 
11754-12705 & $11.1$  & $66.0 \pm 1.7$ (s)  & $21.2 \pm 0.6$ & $44.9^{+8.4}_{-9.9}$  & $131.5 \pm 5.3 $ & $12.2 \pm 1.3 $ & $9.2^{+1.5}_{-1.2}$  & $9.2^{+3.1}_{-2.4}$  & $13.0^{+5.1}_{-3.7}$  & $1.4^{+0.6}_{-0.4}$ \\ 
11872-12702 & $10.9$  & $163.3 \pm 1.7$ (p+s+m)  & $115.0 \pm 4.0$ & $53.3^{+2.6}_{-2.6}$  & $228.4 \pm 5.2 $ & $5.0 \pm 0.5 $ & $9.5^{+1.1}_{-1.0}$  & $13.1^{+2.0}_{-2.2}$  & $14.1^{+3.8}_{-3.2}$  & $1.5^{+0.4}_{-0.4}$ \\ 
11957-9101 & $11.0$  & $4.8 \pm 1.1$ (m)  & $130.7 \pm 20.2$ & $49.7^{+1.4}_{-1.4}$  & $338.0 \pm 20.5 $ & $7.3 \pm 3.5 $ & $8.1^{+3.1}_{-2.3}$  & $26.0^{+7.5}_{-6.7}$  & $11.3^{+4.7}_{-3.4}$  & $1.4^{+0.8}_{-0.6}$ \\ 
11958-6104 & $10.9$  & $77.6 \pm 1.5$ (s+m)  & $89.6 \pm 79.3$ & $26.5^{+4.6}_{-5.6}$  & $403.0 \pm 5.5 $ & $6.6 \pm 1.1 $ & $9.8^{+1.5}_{-1.4}$  & $23.8^{+6.8}_{-5.1}$  & $15.6^{+3.4}_{-2.9}$  & $1.6^{+0.4}_{-0.4}$ \\ 
11963-9102* & $10.4$  & $65.3 \pm 0.9$ (s+m)  & $115.6 \pm 0.8$ & $43.4^{+3.0}_{-3.2}$  & $223.1 \pm 2.1 $ & $5.8 \pm 1.0 $ & $5.1^{+0.8}_{-0.7}$  & $36.8^{+3.4}_{-3.2}$  & $5.8^{+0.6}_{-0.6}$  & $1.1^{+0.2}_{-0.2}$ \\ 
11970-12704 & $10.9$  & $89.9 \pm 1.4$ (p+s)  & $42.7 \pm 5.1$ & $46.5^{+5.3}_{-5.8}$  & $192.6 \pm 5.4 $ & $8.9 \pm 1.5 $ & $7.7^{+1.4}_{-1.2}$  & $20.1^{+5.2}_{-5.1}$  & $9.3^{+3.1}_{-2.1}$  & $1.2^{+0.5}_{-0.3}$ \\ 
11970-6102 & $10.9$  & $168.2 \pm 0.8$ (s+m)  & $13.0 \pm 2.9$ & $33.6^{+7.8}_{-9.9}$  & $367.7 \pm 15.5 $ & $5.2 \pm 0.9 $ & $5.3^{+0.8}_{-0.7}$  & $26.2^{+8.2}_{-4.7}$  & $12.5^{+2.6}_{-2.5}$  & $2.3^{+0.6}_{-0.6}$ \\ 
11977-1902 & $10.7$  & $77.9 \pm 1.2$ (s+m)  & $51.6 \pm 4.5$ & $52.6^{+7.6}_{-8.5}$  & $334.0 \pm 9.2 $ & $4.4 \pm 0.7 $ & $5.6^{+1.3}_{-0.9}$  & $29.8^{+5.8}_{-5.9}$  & $8.7^{+1.7}_{-1.4}$  & $1.5^{+0.5}_{-0.4}$ \\ 
12084-3702 & $10.3$  & $48.5 \pm 1.9$ (s+m)  & $108.7 \pm 4.6$ & $26.2^{+2.2}_{-2.4}$  & $244.1 \pm 3.0 $ & $4.5 \pm 0.5 $ & $3.0^{+0.3}_{-0.3}$  & $27.8^{+12.0}_{-11.4}$  & $7.5^{+2.9}_{-2.0}$  & $2.5^{+1.0}_{-0.7}$ \\ 
12487-9101 & $10.5$  & $179.0 \pm 1.3$ (p+s+m)  & $155.5 \pm 0.8$ & $51.2^{+1.8}_{-1.9}$  & $155.2 \pm 2.4 $ & $3.4 \pm 0.8 $ & $4.0^{+0.8}_{-0.7}$  & $40.9^{+2.7}_{-2.9}$  & $3.4^{+0.6}_{-0.6}$  & $0.8^{+0.2}_{-0.2}$ \\ 
12490-3703 & $10.5$  & $-2.9 \pm 2.8$ (p+s)  & $103.8 \pm 3.0$ & $21.2^{+1.9}_{-2.1}$  & $247.9 \pm 14.5 $ & $5.6 \pm 1.3 $ & $5.1^{+1.1}_{-0.9}$  & $20.9^{+9.4}_{-6.6}$  & $9.2^{+3.4}_{-2.9}$  & $1.8^{+0.8}_{-0.6}$ \\ 
12700-6102 & $10.4$  & $169.3 \pm 3.3$ (p)  & $131.3 \pm 3.0$ & $36.6^{+9.1}_{-11.7}$  & $333.7 \pm 9.5 $ & $3.9 \pm 0.4 $ & $4.1^{+0.5}_{-0.4}$  & $39.4^{+14.4}_{-11.1}$  & $8.3^{+2.5}_{-1.9}$  & $2.0^{+0.7}_{-0.5}$ \\ 
   \hline 
   \caption{ 
   Col. (1): Galaxy MaNGA name. 
   Col. (2): Logarithmic Stellar Mass. 
   Col. (3): Weighted Disc Position Angle (PA that are equally weighted, p:photometric, s:symetric, m:model). 
   Col. (4): Bar photometric position angle. 
   Col. (5): Weighted disc inclination (photometric + model). 
   Col. (6): Disc circular velocity fitted from equation \ref{eq:Bertola} with $\gamma = 1$. 
   Col. (7): Bar radius estimated from the maximum isophote ellipticity method. 
   Col. (8): Deprojected bar radius. 
   Col. (9): Bar pattern speed. 
   Col. (10): Corotation radius. 
   Col. (11): Rotation rate. 
   } 
   \label{tab:master} 
\end{longtable}\end{landscape}



\bsp	
\label{lastpage}
\end{document}